**Chiral-Induced Spin Selectivity Effect in a 1 nm Thin 1,1'-Binaphthyl-2,2'-diyl Hydrogenphosphate Self-Assembled Monolayer on Nickel Oxide**

*Abin Nas Nalakath[#], Christian Pfeiffer[#], Anu Gupta, Franziska Schölzel, Michael Zharnikov, Georgeta Salvan, Ron Naaman, Marc Tornow*, and Peer Kirsch**

Abin Nas Nalakath, Peer Kirsch
Organic Electronics, Technical University of Darmstadt, Darmstadt, Germany
E-mail: peer.kirsch@tu-darmstadt.de

Christian Pfeiffer, Marc Tornow
Molecular Electronics, Technical University of Munich, Garching, Germany
E-mail: tornow@tum.de

Anu Gupta, Ron Naaman
Department of Chemical and Biological Physics, Weizmann Institute of Science, Rehovot, Israel

Franziska Schölzel, Georgeta Salvan
Institute of Physics, Chemnitz University of Technology, Chemnitz, Germany

Franziska Schölzel, Georgeta Salvan
Research Center for Materials, Architectures and Integration of Nanomembranes, Chemnitz University of Technology, Chemnitz, Germany

Michael Zharnikov
Applied Physical Chemistry, Heidelberg University, Heidelberg, Germany

Peer Kirsch
Merck Electronics KGaA, Darmstadt, Germany

Peer Kirsch
Freiburg Materials Research Center (FMF), University of Freiburg, Freiburg, Germany

**Abstract**

The chiral(ity)-induced spin selectivity (CISS) effect describes the observed correlation between the spin of an electron transferred through a molecule and that molecule's chirality. Primary CISS systems are based on self-assembled monolayers (SAMs) of multiple nanometer-long biomolecules exhibiting thiol-anchored helical chirality. For transforming the concept to real application, more robust molecule-substrate systems are required. Phosphonic and phosphoric acid SAMs coupled to metal oxides can provide the necessary robustness. In this work, we report on studies employing the aromatic, axially chiral organophosphoric acid derivative 1,1'-binaphthyl-2,2'diyl hydrogenphosphate (BNP). Grown as a roughly 1 nm thin SAM on top of $NiO_x$/Ni substrates, the system exhibits a high CISS-magnetoresistance (CISS-MR) of 50-80% when measured using magnetic-conductive atomic force microscopy. For biases above 0.5 V, the magnetoresistance curves could be fitted to a minimal Fowler-Nordheim (FN) tunneling model. From this model, we determined that, depending on the molecules' handedness, electrons of a certain spin direction face an effective tunneling barrier, which is either 80 % higher or 40 % lower compared to the barrier for electrons of opposite spin direction. Due to their small size, compatibility with oxide materials, and commercial availability, these molecules are excellent candidates for the realization of novel organic spintronic devices.

[#]**Abin Nas Nalakath and Christian Pfeiffer contributed equally to this work.**

## 1. Introduction

Since the discovery of the chiral(ity)-induced spin selectivity (CISS) in 1999,[1] the effect has been observed in a wide variety of chiral materials and compounds.[2, 3] A standard configuration for detecting CISS consists of a ferromagnetic Ni substrate (often capped by a thin Au layer) and few-nanometer-long thiolated biomolecules, such as double-stranded DNA or polypeptides, deposited as self-assembled monolayers (SAMs).[4–7] It has been reported that the CISS response in α-helical oligopeptides and in double helix DNA increases linearly with the length of the helix.[8] Consequently, several theoretical approaches utilize helical chirality as a basis for modelling CISS.[9–12] However, while indeed many studies have focused on few-nanometer-long α-helices, CISS has been observed in all types of chirality[2], including one-atom thick layers with 2D chirality[13]. Besides continuous efforts towards a better understanding of the fundamental mechanisms of the CISS effect,[2, 14] its potential role in forthcoming spintronic and quantum applications has moved to the foreground recently.[15–17] Electronic devices based on the manipulation of the spin-polarized electrical currents (spintronics) have been in use since the early 2000s,[18] with main applications in the areas of memory and information processing.[19–21] While a memory structure based on a chiral layer has been proposed previously,[22] device architectures competitive to the spintronics state-of-the-art, including magnetic tunnel junctions (MTJs), giant magnetoresistance (GMR) junctions, and related, mature industrial-scale technologies, are yet to be demonstrated. This is partially related to the fact that most CISS-magnetoresistance (CISS-MR) setups involve a noble metal directly deposited on the ferromagnetic substrate, resulting in, e.g., Au/Ni[23] or Au/Co/Au[5] stacks. These stacks, however, are not compatible with many semiconductor fabrication processes, as they contain Au, which is not CMOS-compatible.[24, 25] In addition, the employed helical biomolecule thiols lack sufficient robustness against thermal or oxidative stress. In this context, the ability to utilize short non-helical and even non-bioorganic materials, which would demonstrate a robust CISS signature in transport, and their combination with a substrate stack that avoids the use of Au, appears to be particularly relevant.[16, 26] Hence, in this work, we set out to design and demonstrate such a system, which is both chemically robust and compatible with most microelectronics fabrication processes. It relies on the use of phosphoric molecule-based SAMs, chemically related to organophosphonate SAMs, for which we have shown that binding them to semiconductor and metal oxides is a robust and reliable pathway for making them an integral part of various functional electronic devices.[27–30] The covalent binding mechanism between the acidic OH moiety of the molecules' anchoring group and surface oxides, as well as the charge transport through phosphonic acid molecules, and their ability to serve as robust passivation layers against oxidation, have been extensively studied.[31–35] Consequently, for the chiral SAM, we were looking for a simple compound that shows high chirality, is readily available in both enantiomers, and can bind covalently to metal oxides. Among several potential candidates, suitable compounds in this context are cyclic 1,1'-bi-2-naphthol (binol) derivatives, which are known in liquid crystal technology for their extraordinary ability to induce chirality into nematic phases.[36, 37] Therefore, we selected commercially available binol phosphoric acid (BNP)[38] as the chiral component on Ni substrates. BNP – as illustrated in Figure 1b and less than 1 nm short – manifests the axial chirality of the binol moiety. The two naphthalene arms are rotated against each other by 53.1°.[39] An overview of the deposition process, which involves immersion in the solution, annealing, and subsequent covalent bonding with the metallic oxide surface, is shown in Figure 1c. A detailed description of the SAM deposition procedure is provided in the SI. The architecture of the entire system is illustrated in Figure 1a along with the used magnetic-conductive atomic force microscopy (mc-AFM) measurement setup.

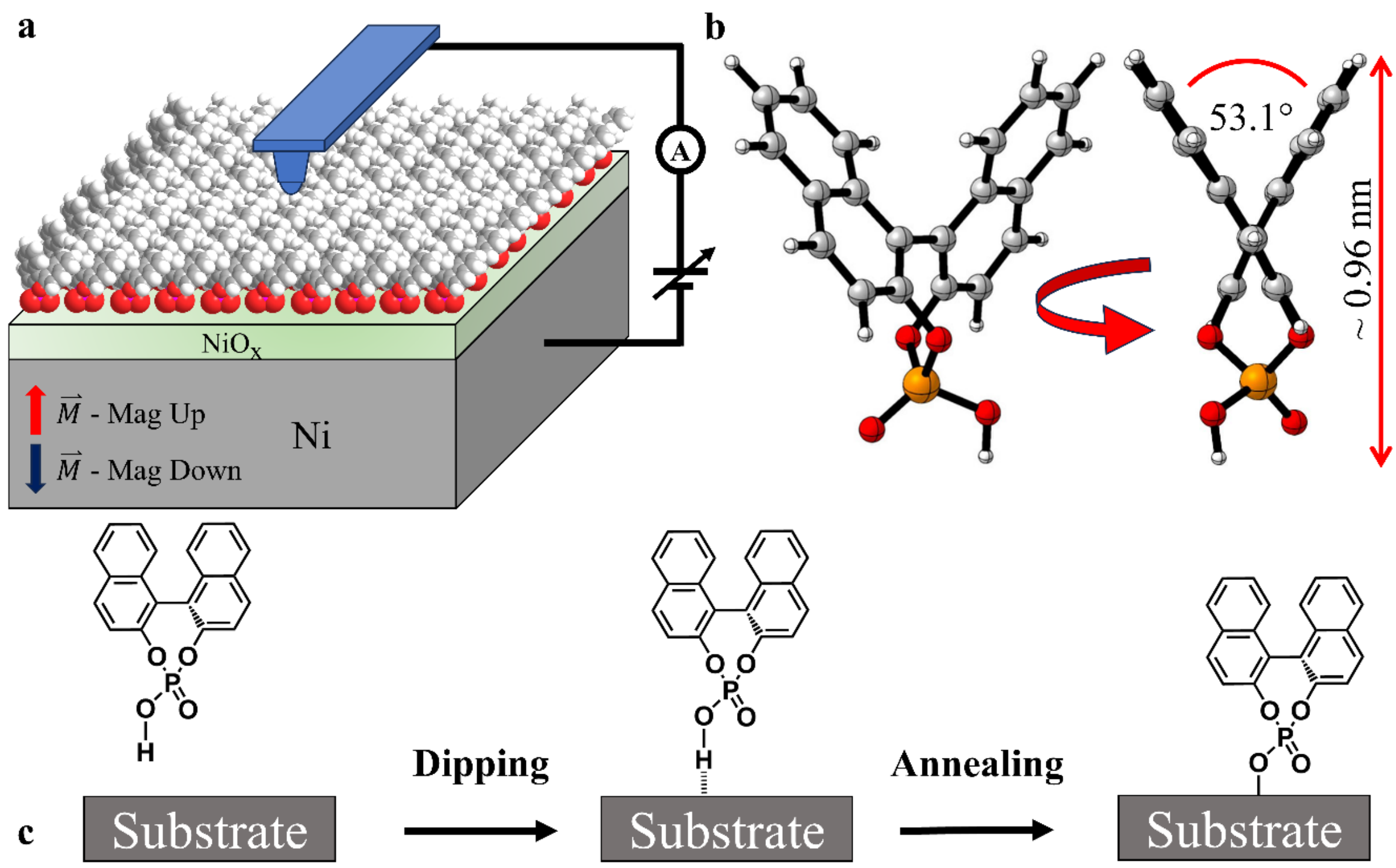

**Figure 1:** (a) Schematic illustration of mc-AFM setup for electrical, magnetic field-dependent CISS-MR measurements. Si, Ti, and Au layers below Ni are omitted. Dimensions are not drawn to scale. Depiction of *R*-BNP within SAM is taken from Chem3D (Revvity, Waltham, USA). (b) Rendering of the *S*-BNP molecule visualized from two different viewing angles.[40] (c) Illustration of the self-assembly process of BNP on $NiO_x$. This process involves a dipping and an annealing step to facilitate a covalent bond formation between the phosphoric acid group and the native metal oxide substrate (only one of several possible binding modes is shown).

We here provide proof of concept for a novel device architecture utilizing the CISS effect, which introduces both a substrate material (metal/metal oxide), which is compatible with most common microelectronics fabrication processes, and an extremely thin synthetic, chiral organic compound monolayer. Not only is the chirality of these molecules preserved upon self-assembly, but mc-AFM measurements indicate strong CISS responses with a CISS-MR > 50%. This finding is remarkable since the substrate material has much lower spin-orbit coupling (SOC) than commonly used Au, which therefore points to the apparently limited influence of the substrate SOC on the extent of the CISS effect. We anticipate our results to provide a viable platform for future CISS-based spintronic applications.

## 2. Results

### 2.1. Surface Analysis

To verify the presence of BNP on $NiO_x$, we conducted X-ray photoelectron spectroscopy (XPS) measurements. As expected, according to the Ni 3p spectra (see Figure S1), the surface of the Ni substrates is oxidized. The thickness of the oxide layer, estimated from the relative weight of the oxide contribution in the entire Ni 3p signal, is 1.3 nm (see the SI, section 2). The oxidation state of Ni is less than in stoichiometric NiO, as emphasized by the lower binding energy (BE) of the Ni $3p_{3/2}$ signal (67.27 eV vs. 69.00 eV for NiO[41]). The C 1s and P 2p spectra of *S*- and *R*-BNP on Ni, individually and as the racemate, are presented in Figures 2a and 2b, respectively. These spectra are referenced against the spectra of *n*-tetradecyl phosphonic acid (C14) on Ni and *n*-hexadecanethiolate (C16) SAM on Au(111) (C 1s only). The latter was prepared by a standard procedure[42], features well-defined properties[43], and serves as an established reference for SAM studies.[44] All carbon spectra in Figure 2a exhibit a single peak at either ~284.7 eV or 284.96 eV corresponding to binol and alkyl backbones, respectively, and differing from the signal of carbon contamination for the blank Ni, recorded at 285.7 eV (we assume that most of this contamination was wiped off upon the SAM formation). Whereas the peaks for C14/Ni look symmetric, those of the BNP films are accompanied by slight shoulders (residual contamination and, probably, the carbon atoms bound to the anchoring group), especially pronounced in the case of *rac*-BNP. Significantly, the intensities of the C 1s signal for the films on Ni are lower than that of the C16 thiol-SAM, which suggests that all former films are monolayers. The numerical evaluation of the C 1s data, using the standard expression for the self-attenuation of the photoemission signal,[45] literature values for the attenuation lengths in SAM-like films,[46] and the thickness of C16/Au (1.89 nm)[43] as a reference, gives the thicknesses of the C14 and BNP films (hydrocarbon matrix only) of ~1.35 nm and ~0.92 nm, respectively. These values correlate well with the lengths of the molecular backbones, which are ~1.8 nm (calculated based on length per methylene group[47]) and 0.96 nm, respectively. The P 2p spectra of all films on Ni in Figure 2b exhibit a single P $2p_{3/2,1/2}$ doublet at a binding energy of ~133.5 eV (P $2p_{3/2}$), corresponding to the anchoring phosphoric acid groups.

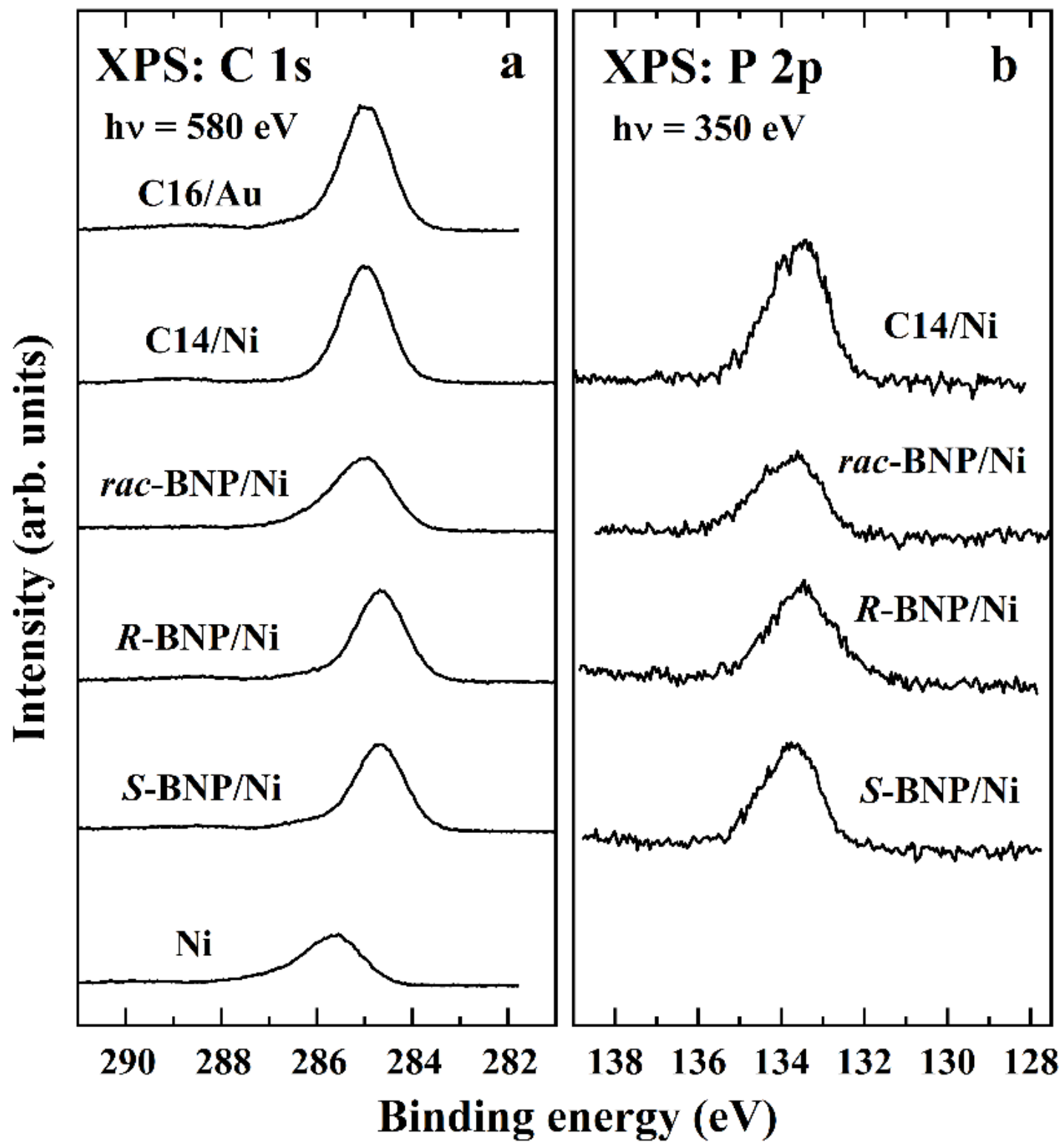

**Figure 2:** C 1s (a) and P 2p (b) XPS spectra of *S*- and *R*-BNP on Ni, individually and as a racemate (*rac*), along with the reference spectra of C14/Ni, C16/Au, and bare Ni.

However, these data do not allow us to derive information about the exact binding mode of these groups to the substrate[48], which can generally involve a variety of different motifs, from monodentate to tridentate.[33, 49] Yet, we can estimate the packing density of the BNP monolayers based on these data, assuming that the packing density of C14/Ni is similar to that of C16/Au ($4.63 \cdot 10^{14}$ molecules·cm$^{-2}$).[43] Calculating the intensities of the P 2p signals and taking into account the attenuation by the binol/alkyl matrix, we got a packing density of $\sim 1.9 \cdot 10^{14}$ molecules·cm$^{-2}$ for the *R*-BNP and *S*-BNP monolayers and ~$1.65 \cdot 10^{14}$ molecules·cm$^{-2}$ for the *rac*-BNP film, which is in good agreement with the layer structure in *R*- and *S*-BNP crystals featuring $1.817 \cdot 10^{14}$ molecules· cm$^{-2}$ (see also SI).[39] Thus, the packing density of the racemate film is ~20% lower than that of the enantiomer monolayers. This could be due to the triclinic space group crystallization known to take place in *rac*-BNP[50], which stands in contrast to the layered crystalline structures that occur in enantiopure BNP[39].

Complementary information about the properties of the BNP films is provided by near-edge X-ray absorption fine structure (NEXAFS) data (see Figures S2 and S3), using the established approach.[51] Here, the BNP samples show the characteristic absorption features of naphthalene, which is the major building block of binol. The small but distinct dichroism in the spectra of the enantiomer SAMs suggests a certain degree of orientational order and upright molecular orientation. In contrast, the lack of any dichroism in the *rac*-BNP spectra indicates a more disordered monolayer compared to the enantiomer samples.

The surface roughness analysis of bare and BNP-coated samples, as determined using AFM, is presented in Figure S4. Furthermore, AFM-based scratching experiments were conducted, in which the applied force of an AFM tip was calibrated to mechanically remove the soft organic monolayer in contact mode. Figure 3d depicts a 3×3 $\mu m^2$ tapping mode image of a scratched *S*-BNP-coated sample, indicating successful removal of the organic layer in the central square. Height variations along the horizontal axis of the image are averaged over the scratched area and shown in Figure 3e. Calculated as the height difference between the scratched and unscratched parts in the profiles, the *S*-BNP layer is determined to be 0.6 nm thin. Additional height analysis for samples with *R*-BNP and *rac*-BNP is shown in Figure S5. It yields monolayer thickness values of 0.5 nm, 0.6 nm, and 0.8 nm for *rac*-BNP, *S*-BNP, and *R*-BNP, respectively. Grazing incidence X-ray reflectivity (XRR) measurements were performed at different positions to evaluate the structural uniformity of the films. In Figure 3b, a representative XRR recording is shown with its corresponding simulated fits. The simulated profile closely matches the experimental Kiessig fringes, confirming the accuracy of our fitting procedure. The fit is not only able to reproduce the thicknesses specified for the substrate wafer but also determines a NiO thickness of 1.33 nm and a SAM thickness of 0.98 nm, which is in good agreement with XPS and AFM data (see Tables S1, S2, and S3 for comparison).

## 2.2. Circular Dichroism

The circular dichroism (CD) spectra for the BNP molecules in ethanol solution are presented in Figure 3a. They show mirror-image Cotton effects for the enantiomers, while the racemic mixture is optically inactive. The degree to which the chirality is expressed optically can be quantified by the dissymmetry factor $g$ [52, 53], defined as $g = \Delta\varepsilon/\varepsilon$ , where $\Delta\varepsilon$ is the difference in the molar absorption coefficient for left and right circularly polarized light and $\varepsilon$ is the molar absorption coefficient of the molecule.[54] Recent work on the twisted-acene systems shows that the sign and magnitude of the CD peaks (associated with $g$) qualitatively track the sign and magnitude of the CISS-derived spin polarization, suggesting that the factor $g$ captures the chiral electronic response which underlies spin selectivity. [55] Notably, while the parameters $\varepsilon$ and $\Delta\varepsilon$ scale with the number of naphthyl chromophores, the dissymmetry factor offers a normalized measurement of chirality. For the *R*-BNP enantiomer, there is a positive band at 215.5±0.1 nm ($\theta \approx -150$ mdeg, $g \approx -1.6 \cdot 10^{-3}$), followed by a negative band at 227.5±0.1 nm ($\theta \approx 191$ mdeg, $g \approx 2.5 \cdot 10^{-3}$). *S*-BNP exhibits the practically exact mirror-image response. The close alignment of the absorption maxima and the mirror-image CD signals support the conclusion that the observed optical activity results from exciton coupling between the naphthyl transition dipoles. The magnitudes of $g$ ($\sim 10^{-3}$) align with those reported for similar binaphthyl frameworks[56–58], highlighting the strong intrinsic chirality of BNP. This behavior is consistent with findings by Amsallem *et al.*, which demonstrate that thiolated binaphthalene and ternaphthalene exhibit different $\varepsilon$ and $\Delta\varepsilon$ values, yet have similar $g$ factors that correlate with comparable spin polarization efficiencies.[56] Additionally, the relationship between BNP's absorption bands and its mirror-image CD signals matches the behavior observed in other binaphthyl derivatives during light-driven desorption experiments.[59]

Importantly, we also recorded CD spectra of *R*-BNP, *S*-BNP, and *rac*-BNP in SAMs on $NiO_x$/Ni/sapphire. The CD measurements show the chirality of the self-assembled monolayer on the surface and confirm that the BNP SAM preserves its chiral nature after adsorption. As shown in Figure 3c, the overall spectral shape is preserved when probed in monolayer form, consistent with the solution CD spectra (Figure 3a), and the enantiomers continue to exhibit roughly mirror-image CD bands in the 220-260 nm region.

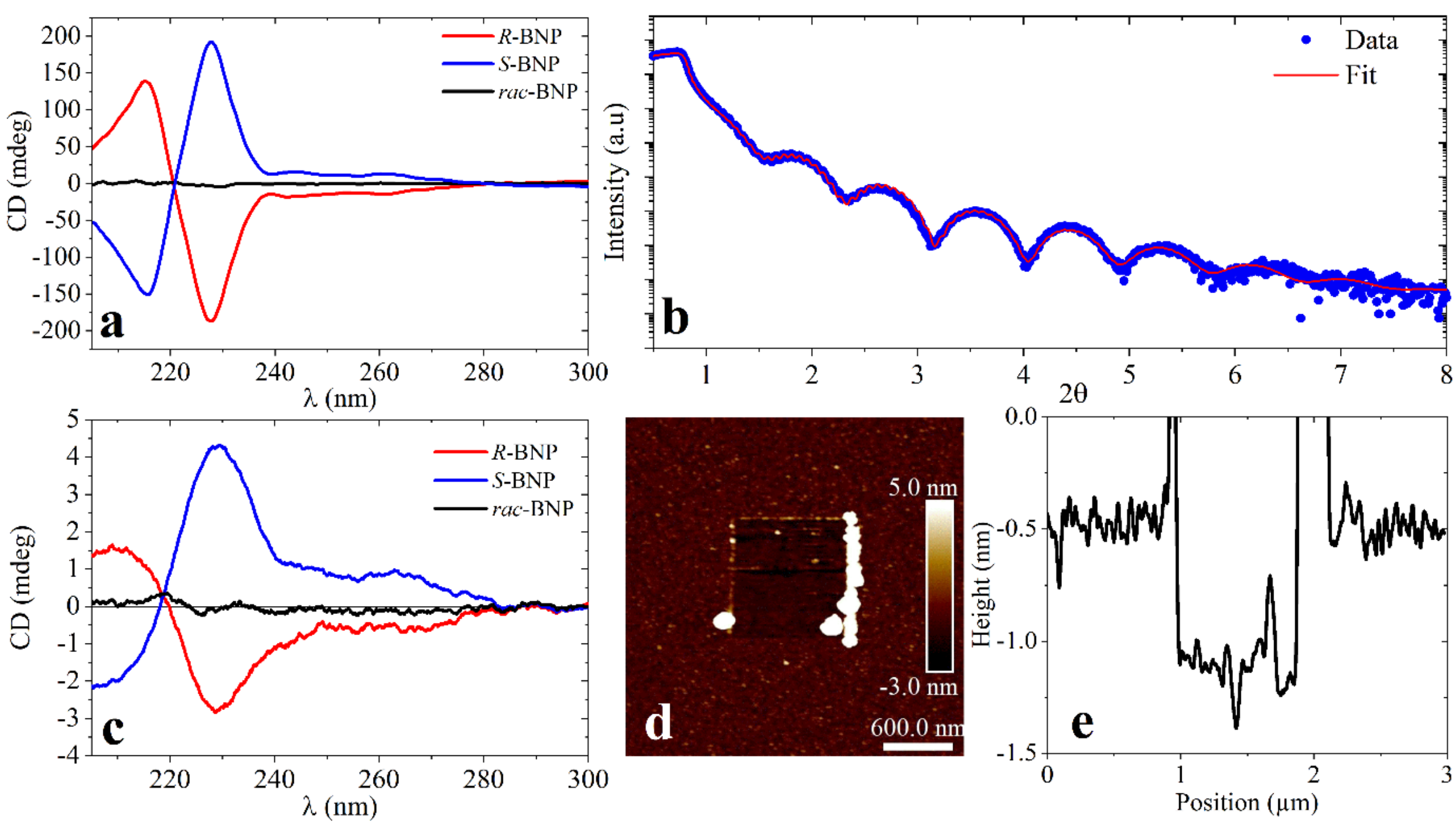

**Figure 3:** Characterization of the BNP molecules and the respective monolayers on $NiO_x$: (a) CD spectra of *R*-BNP (red), *S*-BNP (blue), and *rac*-BNP (black) in ethanol (0.01 mM solution). (b) XRR results for *rac*-BNP grown as a SAM on $NiO_x$/Ni substrate. Measured reflected intensity is drawn in blue circular markers over the angle of incidence 2θ. The red line indicates the fit. (c) CD spectra of *R*-BNP (red), *S*-BNP (blue), and *rac*-BNP (black) monolayers grown on $NiO_x$/Ni/sapphire. (d) 3×3 $\mu m^2$ AFM tapping recording of a scratched sample of an *S*-BNP monolayer. (e) Mean height profile along the horizontal direction over the scratched area in (d).

The CD spectra of the thin films showed maximum ellipticities of $\pm 3 - 4$ mdeg, that were weaker than those observed in solution (Figure 3c). Additionally, the spectral bands of the thin films were significantly broader than those of the corresponding CD spectra in solution. The $g$-factors of the thin films were determined to be $5.0 \cdot 10^{-5}$ for *S*-BNP and $5.2 \cdot 10^{-5}$ for *R*-BNP. The reduction in the g-factor for the thin films can be attributed to the experimental configuration, which includes both the substrate and the SAM. The nickel substrate shows strong absorption and reflection in the UV region.[60] As in the calculation of the g-factors of the thin films, reflection is not taken into account, the genuine g-factors of the thin films might in fact be larger than the experimentally estimated value. Both enantiomeric thin films exhibited similar trends in their solution CD spectra. In contrast, the racemic films showed no significantly detectable CD signal, indicating the absence of overall chirality in the system, consistent with the solution-state CD results.

### 2.3. CISS-Magnetoresistance

Figures 4a and b show the mc-AFM results for the *S*-BNP and *R*-BNP SAMs, respectively. As is visible, the current magnitude depends on the magnetization of the underlying Ni layer. While the current passing through downward magnetized Ni seems to be favored by *S*-BNP (Fig. 4a), *R*-BNP (Fig. 4b) exhibits the opposite behavior. For samples coated with *rac*-BNP, the measured magnetoresistance appears to be statistically insignificant, with overlapping error bars (Fig. 4c). The individual IV-curves measured for *S*-BNP, *R*-BNP, and *rac*-BNP are provided in Figures S8, S9, and S10, respectively. It is noteworthy that, while the results in Figure 4 are based on statistics of multiple positions, the measurements on each individual position also exhibit the expected CISS response. The measurements taken for *rac*-BNP (Fig. S10) do show small CISS-MR effects on individual positions but appear insignificant when averaged. We attribute this to a local excess of one enantiomer for individual mc-AFM junctions. As is visible in Figures S8-S10, several IV measurements reach the measurement system's maximum limit of 10 nA. This can lead to an underestimation of the mean current for applied bias above 0.5 V. Using the currents measured when the external magnetic field is pointed upward, here denoted as $I_{\uparrow}$ and downward $I_{\downarrow}$, respectively, *CISS-MR* in per cent can be calculated according to

$$CISS\text{-}MR = \frac{I_{\uparrow} - I_{\downarrow}}{I_{\uparrow} + I_{\downarrow}} \cdot 100. \quad (1)$$

The resulting CISS-MR is plotted in Figure 4d for *R*-BNP (green), *S*-BNP (black), and *rac*-BNP (magenta) for absolute voltages between 0.6 V and 1 V. This yields a CISS-MR for *R*-BNP of $\approx 81 \pm 5\%$, for *S*-BNP of $\approx -51 \pm 8\%$, and for *rac*-BNP of $\approx -6 \pm 8\%$. These values represent the average and standard deviation of the respective data points shown in Figure 4d. The three results for the CISS-MR, as calculated over the entire voltage range, are provided in Figure S11.

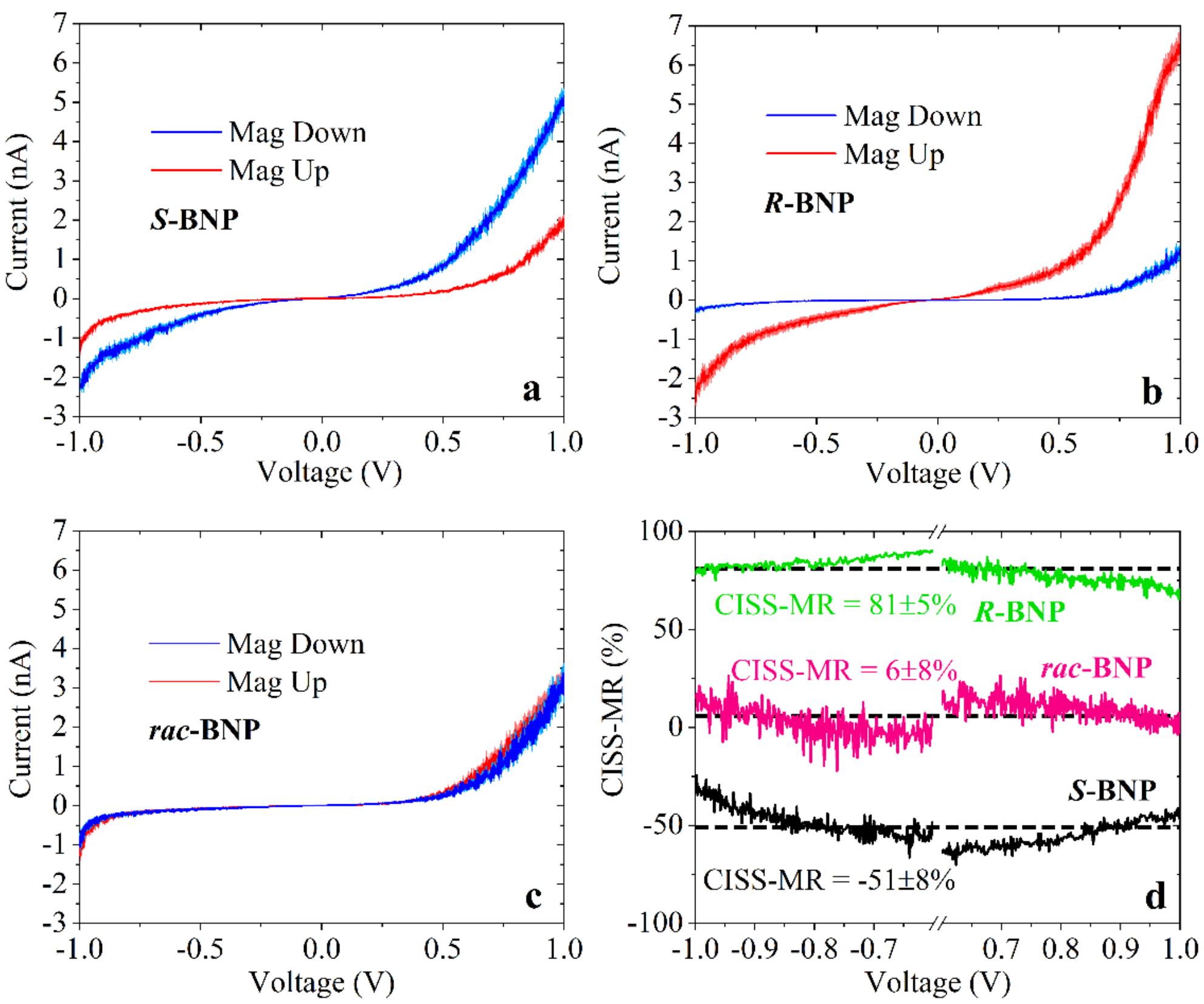

**Figure 4:** Magnetic field-dependent mc-AFM measurements of the *S*-BNP (a), *R*-BNP (b), and *rac*-BNP (c) SAMs. Blue (red) lines represent average IV curves when the magnetic field (0.5 T) is pointed downwards (upwards), perpendicular to the sample plane. Solid lines indicate the mean curves. Shaded, lighter areas indicate the standard error of the mean. (d) Determined CISS-MR for *R*-BNP (green), *S*-BNP (black), and *rac*-BNP (magenta) for voltages between −1 V and −0.6 V, and between +0.6 V and +1 V.

## 3. Discussion

The measured CISS-MR is roughly in the range of values for thiol-functionalized binaphthyl monolayers[56], underscoring the robustness of axial chirality in inducing spin selectivity. If we consider the thickness of the BNP SAM alone, and the previous observation that CISS increases with the length of a helical structure[8, 61, 62], the achieved MR is remarkably high. Notably, the measured magnitude falls within the range of comparable values for roughly 12 nm long dsDNA and exceeds that for oligopeptides of > 2.5 nm length.[8]

A unique characteristic of the BNP system is the intrinsically defined orientation of the individual molecules on the surface: BNP has basically no conformational degrees of freedom, and the binding to the substrate can occur only through the phosphate moiety. Therefore, it can be expected that the chiral surface looks actually very similar to the layers observed in the crystal structures of *R*- and *S*-BNP (see Fig. S6).[39] BNP has an approximate $C_2$ symmetry with a continuous symmetry measure (CSM) $S(C_2)$ of 0.9309 and a continuous chirality measure (CCM) of 8.3447.[63] On binding to the substrate, the $C_2$ axis of BNP is ideally oriented perpendicularly to the substrate plane.

As noted above, the molecules are grown onto the native oxide of Ni, which is, in principle, expected to be NiO with at least partial hydroxyl termination.[64, 65] Therefore, the spin-polarized current injected from the ferromagnet must pass an oxide barrier. Bulk stoichiometric NiO has a wide band gap (~ 3.6 - 4.0 eV)[66–68] and acts as a Mott insulator.[69] If this applies equally to the native thin film oxide, the electrons would first have to pass through a potential barrier before entering the chiral layer, resulting in a TMR[70, 71] effect instead of a GMR[72, 73] effect. XPS data indicate that this native oxide exists in a slightly sub-stoichiometric oxidation state. It has been previously shown that in $NiO_x$, a lower oxidation state can result in higher resistance.[74, 75] However, especially in ultrathin films of ordered NiO, smaller effective band gaps and stronger coupling to the metal have been reported.[76, 77] Further, a narrower effective gap can also be correlated with a lower level of ordering and non-ideal stoichiometry[78], which is expected for the natively grown oxide film, such as on the present substrates. We therefore conclude that the thin layer between the ferromagnetic metal and the chiral layer may be significantly less insulating than bulk NiO. This would mean that the measured difference in resistance is closer to a GMR than to a TMR.

Nickel oxide is known to have antiferromagnetic characteristics at room temperature.[79–81] Generally, this could lead to a depolarization of the current injected into the SAM due to the short spin decay lengths in antiferromagnets.[82] However, assuming that spin-polarized current is governing the CISS effect, the measured MR does not indicate any strong antiferromagnetic properties dominating the charge transport through the $NiO_x$. This could be due to the strong dependence of the Néel-temperature on the layer thickness. For example, a five-monolayer-thick film of epitaxially grown NiO on MgO has a reduced Néel-temperature of 295 K.[83] Altieri *et al.* reported a Néel temperature below 40 K for three monolayers of NiO grown on MgO and a comparatively high temperature of 390 K when grown on Ag.[84]

As pointed out previously, the measure of *CISS-MR* (also sometimes referred to as spin polarization), as calculated in equation (1), does not refer to the spin polarization induced by the chiral molecule alone[85], but rather reflects the differences in resistance of the entire stack based on the direction of the magnetic field. Notably, the CISS effect has also been observed in one-atom-thick molecular layers with 2D chirality[13] or in TMR junctions with chiral molecules[86]. In the latter study, the CISS was explained by a spin-polarized or spin-selective interface effect. Importantly, in many cases, the IV curves are anti-symmetric with respect to the bias voltage, i.e., *I*(+*V*) is equal to -*I*(-*V*).

The observed non-anti-symmetry in our measurements may stem from the different work functions and electronic structures of the materials in our stack: Ni (5.15 eV)[87], NiO (band gap ≈ 4.3 eV [88], work function ≈ 5.2 – 6.7 eV [65]), and Pt (5.65 eV)[87]. Rikken and Avarvari commented on the role of device asymmetry in CISS-MR results, pointing out that while the electrode asymmetry appears to be a prerequisite for CISS-MR, results typically still report antisymmetric IV curves, possibly due to the effect being below the detection limit of common mc-AFM techniques.[89] In fact, to a certain extent diode-like IV curves under a magnetic field are generally associated with the electrical magnetochiral anisotropy (eMChA) effect, [89, 90] indicating that it may have contributed to the results in Figure 4. However, the results presented here show that the polarity of the current rectification is independent of the applied magnetic field, suggesting that eMChA does not significantly affect the observations above.

In many nanometer-thin SAMs, coherent tunneling has been reported as the predominant charge transport mechanism.[91–93] Here, electrons are expected to directly traverse a rectangular barrier (at low bias) in a single step. The characteristic feature of this model is exponential current attenuation with distance according to the relation for the conductance $G$

$$G = G_C \cdot \exp(-\beta d), \tag{2}$$

where $G_C$ is a reference conductance, namely the zero-length (contact) conductance, $d$ is the thickness of the tunnel barrier, and $\beta$ is the junction-characteristic current decay coefficient.[94] Exemplary, by systematically measuring alkanethiols of varying lengths, a mono-exponential current attenuation was observed with a decay constant of $\beta \approx 0.5 - 1\ \text{Å}^{-1}$.[95]

The simplest phenomenological model to assess CISS would also see the chiral surface as a single rectangular tunnel barrier.[14] While the length $d$ remains the same for both enantiomers, and in our case, covers the sum of both the oxide barrier and the SAM thickness, the height of the barrier appears different for electrons of opposite spins, which is indicated in Figure 5a. The use of a varying barrier height model to evaluate CISS is well documented in the literature over different material systems and especially in the context of interface effects.[90, 96–100] In the presented experimental setup, we cannot evaluate $\beta$ as there is no variation of $d$. Further, a fitting for a low-bias Simmons-type model[101] would require many critical assumptions, including the exact area of the junction. However, with increasing applied bias, the shape of the tunneling barrier can change, potentially reaching a Fowler-Nordheim (FN) regime, in which the electron tunnels through a triangularly shaped barrier.[94, 102] This is illustrated in Figure 5b. In a reduced FN model, the current density $J$ can then be determined as[103]

$$J = \alpha E^2 \cdot \exp\left(-\frac{B}{E}\right) \tag{3}$$

with

$$\alpha = \frac{m_e}{m^*}\frac{q^3}{8\pi h \phi_{\mathrm{B}}} \tag{4}$$

and

$$B = \frac{8\pi}{3}\left(2\frac{m^*}{h^2}\right)^{1/2}\frac{\phi_{\mathrm{B}}^{3/2}}{q}. \tag{5}$$

Here, $E$ is the electrical field, $J$ the current density, $\phi_{\mathrm{B}}$ the effective barrier height, $q$ the electron charge, $m_e$ the electron mass, $m^*$ the effective electron mass in the barrier materials, and $h$ the Planck constant.[103]

As can be seen in the FN-plot in Figure 5c, for an *R*-BNP-coated sample, a clear FN-like regime can be observed. Figure S12 provides the FN plots for both enantiomers and the racemic mixture, showing that, for all samples, the charge transport indeed exhibits an apparent FN regime above 0.5 V (2 $V^{-1}$). The observed linear decrease of $\ln(I/V^2)$ indicates that FN tunneling may well be the primary transport mechanism in that voltage range. Based on this FN tunneling model, we estimate the ratio of the effective tunneling barriers

$$\frac{\phi_{\mathrm{B}\uparrow}}{\phi_{\mathrm{B}\downarrow}} = \left(\frac{s_\uparrow}{s_\downarrow}\right)^{2/3}, \tag{6}$$

where $\phi_{\mathrm{B}\uparrow}$ and $\phi_{\mathrm{B}\downarrow}$ are the effective barrier heights (rectangular, at zero electric field) calculated for currents $I_\uparrow$ and $I_\downarrow$, respectively. The parameters $s_\uparrow$ and $s_\downarrow$ represent the slopes of the linear fits to the data within the FN regime, by using the FN relation in equation (3).

The fitting is based on a minimal model that is described in detail in the SI. It yields that for *R*-BNP the ratio of the effective barrier heights is $\frac{\phi_{\mathrm{B}\uparrow}}{\phi_{\mathrm{B}\downarrow}} = 0.6$, whereas for *S*-BNP it is $\frac{\phi_{\mathrm{B}\uparrow}}{\phi_{\mathrm{B}\downarrow}} = 1.8$. In other words, for *S*-BNP, electrons with spin up experience an 80% larger tunnel barrier than electrons with spin down. As expected, the ratio is nearly 1 for the racemic mixture. The thereby derived barrier differences are $|\Delta\phi_{\mathrm{B}}| \approx 90$ meV for *S*-BNP and $|\Delta\phi_{\mathrm{B}}| \approx 130$ meV for *R*-BNP, and a mere $|\Delta\phi_{\mathrm{B}}| \approx 20$ meV difference for rac-BNP (see Fig. 5, Table S4, and the SI, section 9 for the derivation). We want to emphasize that this determination of absolute barrier differences is based on several simplifying assumptions, such as an effective electron mass of $m^* \approx 0.5\, m_e$ (based on reference value in silicon oxide barrier[104]), that the tunnel barrier is made up of both the oxide (~ 1.3 nm) and the BNP SAM (~ 1 nm), and that the electric field is uniformly distributed over the entire barrier. Furthermore, the model does not aim at explaining the physical mechanisms of the CISS effect. Notwithstanding that, the above (phenomenological) effective barrier height ratio $\frac{\phi_{B\uparrow}}{\phi_{B\downarrow}}$ remains a reliable figure and we propose it as a quantitative metric, possibly also in future theoretical work.

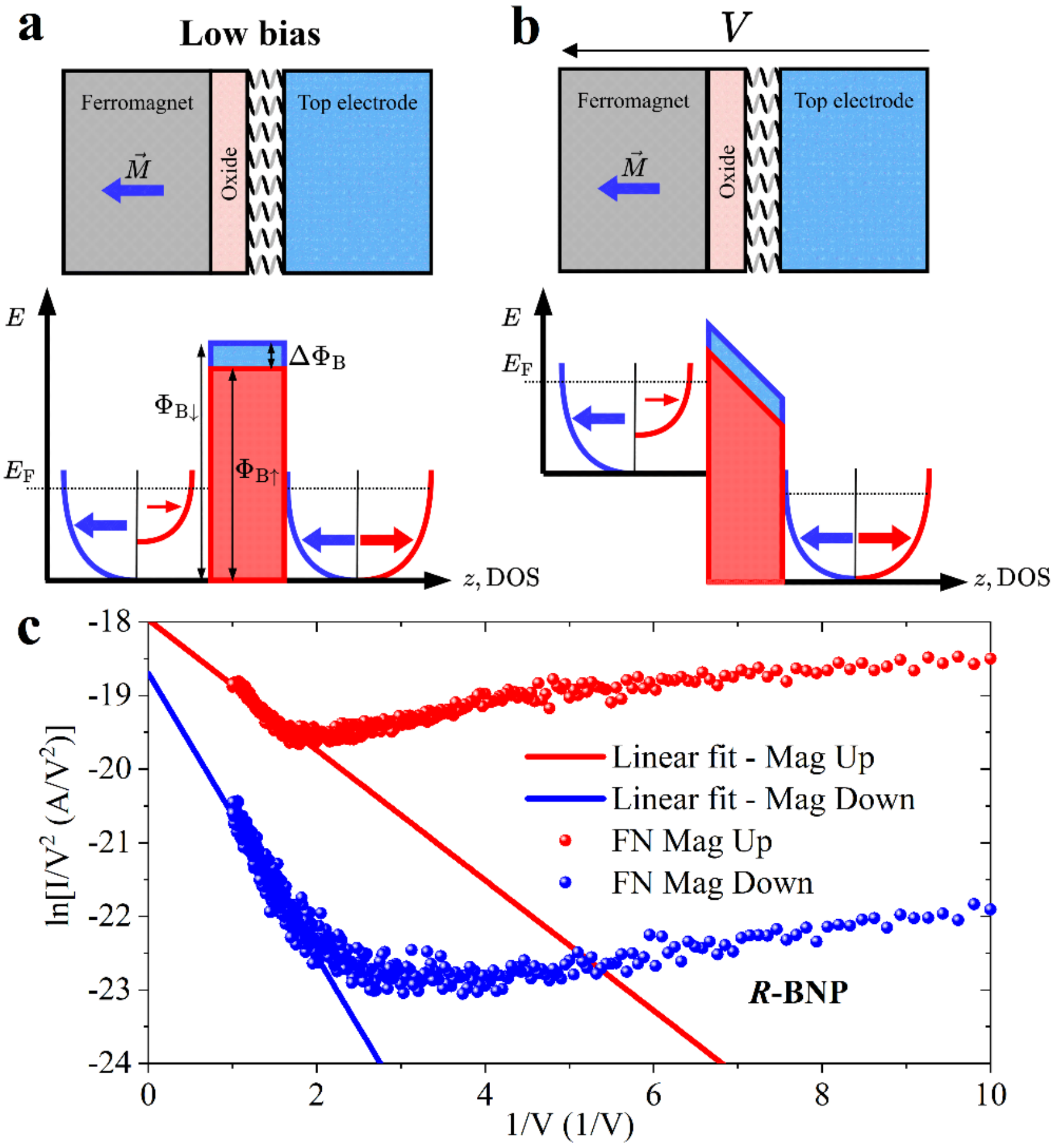

**Figure 5:** Simplified tunneling models, adapted to CISS: (a) Schematic illustration of phenomenological direct tunneling model at low bias. Case for downwards pointing magnetization and *R*-chirality is shown. Parabolas indicate the densities of states (DOS) with spin down (blue) and spin up (red) in the ferromagnet and top electrode. Due to zero or low bias, the Fermi levels of the ferromagnet and top electrode are aligned. The tunneling barrier is made up of oxide and SAM. In this reduced model, electrons with spin down face a higher barrier ($\phi_{B\downarrow}$, blue) than electrons with spin up ($\phi_{B\uparrow}$, red). The difference in barrier height is denoted as $\Delta\phi_B$. (b) Schematic illustration of the phenomenological FN tunneling model. The triangular FN tunneling barrier is indicated to be spin-dependent. Due to applied bias, the Fermi levels are no longer aligned. (c) FN plot for mc-AFM mean current for *R*-BNP. Scatter plot indicates data points. Solid lines represent linear fits in the region [1 $V^{-1}$, 2 $V^{-1}$], but are drawn

over a larger x-axis range for better visibility. Blue (red) color refers to currents when the magnetic field is pointing downwards (upwards).

## 4. Conclusions

To summarize, we demonstrated a spin valve architecture based on a BNP/$NiO_x$/Ni stack. The organic monolayer, which comprises the chiral phosphoric acid molecules anchored to the oxide surface, was shown to be roughly 1 nm thin and to have a packing density in the range of $1.6 \cdot 10^{14}$ $\mathrm{cm}^{-2}$ - $1.9 \cdot 10^{14} \mathrm{cm}^{-2}$ for both enantiomers. The CD response indicates that the chirality is preserved upon self-assembly. The magnetic-electrical characterization shows a clearly observable CISS-based resistance change with a CISS-MR of ≈50 - 80 %. We assign this recorded difference in resistance to the (anti)parallel spin polarization in both the magnetic layer and the chiral tunneling barrier. We believe the spin-selective transport in the latter to be caused by the CISS effect resulting from the axial chirality of BNP. Moreover, we observe a clear fingerprint of FN tunneling as the predominant charge transport mechanism at biases above 0.5 V. The ratio of both effective FN tunneling barriers, each corresponding to either injected spin direction, was derived. This indicates that, phenomenologically, spin-up electrons face a roughly 80 % higher barrier than spin-down electrons when facing an *S*-BNP-based oxide/SAM barrier, whereas the barrier in the same scenario appears to be 40 % lower for the *R*-BNP case. We propose this ratio as a viable measure for the CISS response of a chiral monolayer or interface. It may serve to benchmark different CISS platforms against each other and possibly help further decode the contribution of interface effects on the overall CISS response.

Our presented spin valve architecture demonstrates for the first time that a 1 nm thin SAM of axially chiral organophosphate deposited on a metal oxide, can serve as a basis to a robust CISS device. This result was achieved despite the absence of a helical structure, while at the same time using a substrate with low SOC. Utilizing simple, commercially available binol-based phosphoric acids, this comprises an important step towards the adoption of CISS-based devices by the electronics industry. The fabrication of a full solid state CISS device using a chiral SAM is part of an ongoing project and will be subject of a future communication.

## 5. Experimental Section

*Sample preparation:* Substrates were provided by SIEGERT WAFER GmbH (Aachen, Germany): Ni (100 nm) / Ti (10 nm) / $p^{++}$-Si (525 μm) / Ti (10 nm) / Au (100 nm). Before SAM growth, samples were treated either with plasma (40 W, 0.3 mbar, 30 s) or UV-ozone (4 W, 254 nm). For SAM growth, samples were immersed in 10 mM of BNP dissolved in THF or ethanol for 72 h, then annealed at 80 °C for 1 h, rinsed with isopropanol, and again annealed at 80 °C for 10 min.

*Atomic Force Microscopy:* Tapping and scratching images were taken with a Dimension V AFM (Bruker/Veeco, Billerica, Massachusetts, United States) and the diamond-like-carbon-coated tip 190DLC (BudgetSensors, Sofia, Bulgaria).
IV-curves were taken with a custom-built mc-AFM with an electromagnet and a Beetle Ambient AFM. As conductive tips, Pt-coated DPE-XSC11 (MikroMasch, Sofia, Bulgaria) were utilized. For magnetic field-dependent I-V measurements, a (constant) tip-surface contact force of 8-10 nN was applied.

*X-ray Photoelectron Spectroscopy:* XPS measurements were conducted at the bending magnet HE-SGM beamline of the synchrotron storage ring BESSY II in Berlin. Experiments were done at room temperature in ultra-high vacuum (base pressure: $\approx 1\times10^{-9}$ mbar). The binding energy scale is referenced to the Au $4f_{7/2}$ at 84.0 eV.[105]

*Near-edge X-ray Absorption Fine Structure Spectroscopy:* The NEXAFS spectra were collected at the same beamline as the XPS data. The spectra were recorded at the C K-edge in the partial electron yield mode with a retarding voltage of −150 V. The incidence angle of the primary X-rays was varied. The energy resolution was ~ 0.3 eV. The photon energy scale was referenced to the pronounced π* resonance of HOPG at 285.38 eV.[106]

*Circular Dichroism:* Thin films were measured on SAM/Ni(30 nm)/Sapphire using a JASCO J-1500 CD spectrophotometer (JASCO Corporation, Tokyo, Japan). Results were collected in the range 205 – 500 nm with 200 $nm\cdot min^{-1}$ scan speed, data pitch of 0.1 nm, digital integration time of 2 s, and a 1 nm bandwidth. Measurements of BNP solutions (0.01 mM in ethanol) were done in a quartz cuvette (path length 1 cm) over the same wavelength with 1000 $nm\cdot min^{-1}$ scan speed, data pitch of 0.5 nm, digital integration time of 0.125 s, and a 1 nm bandwidth.

*X-ray Reflectivity:* The SmartLab diffractometer (Rigaku Corporation, Tokyo, Japan) was used with monochromatic Cu $K\alpha_1$ radiation, with a divergent slit with parallel-beam optics and a 2 mm Soller slit. Fit was done using the GenX software.

More details on all methods are provided in the SI.

**Acknowledgements**

C.P. and M.T. thank the ZEITlab (Technical University of Munich) and its technical staff: R. Emling, A. Kwiatkowski, R. Mittermeier, P. Röthlein, and S. Ulu. M.Z. thanks the Helmholtz Zentrum Berlin (HZB) for the allocation of synchrotron radiation beamtime at BESSY II and financial support, as well as Dr. M. Brzhezinskaya (HZB) for her assistance during the experiments at the synchrotron. A.N. and P.K. would like to express their gratitude to Prof. Dr. L. Alff, Prof. Dr. R. Stark and their research groups for granting access to the laboratory facilities, especially the X-ray diffraction (XRD) and AFM instrument. The authors also appreciate Prof. Dr. M. Reggelin for his generosity in allowing access to the circular dichroism (CD) spectroscopy setup. F.S. and G.S. gratefully acknowledge funding through German Research Foundation (DFG), TRR-386, TP B3, project number 514664767.

**Author contributions**

The concept of the study was developed jointly by A.N., C.P., M.T., and P.K.. They also prepared the manuscript. A.N. and C.P. fabricated the samples and carried out AFM, XRR, SQUID measurements. A.G. and R.N. provided the mc-AFM measurements. M.Z. conducted XPS and NEXAFS experiments and provided the interpretation of the results. F.S. and G.S. carried out CD experiments. A.N., C.P., A.G., M.Z., and F.S. contributed to the data analysis.

**Data Availability Statement**

The data that support the findings of this study are available from the corresponding author upon reasonable request.

**Conflict of Interest**

The authors declare no conflict of interest.

# Supporting Information

**Chiral-Induced Spin Selectivity Effect in a 1 nm Thin 1,1'-Binaphthyl-2,2'-diyl Hydrogenphosphate Self-Assembled Monolayer on Nickel Oxide**

*Abin Nas Nalakath[#], Christian Pfeiffer[#], Anu Gupta, Franziska Schölzel, Michael Zharnikov, Georgeta Salvan, Ron Naaman, Marc Tornow*, and Peer Kirsch**

## 1. Sample Preparation

Samples are based on substrates supplied by SIEGERT WAFER GmbH (Aachen, Germany) and utilize a double-side polished 100 mm highly p-doped (boron, $\rho < 0.005\ \Omega$cm) silicon wafer. 10 nm Ti were sputtered on back-and front side as adhesion layers. Subsequently, 100 nm Au and 100 nm Ni were sputtered on the back- and front side, respectively. Native oxides in between deposition steps were removed via ion beam polishing. The $8 \times 8\ mm^2$ samples were cleaned via acetone and isopropanol sonication. For all (mc-)AFM measurements, the surfaces were oxygen plasma treated at 80 W and 0.3 mbar for 30 s (Smart Plasma, plasma technology GmbH, Herrenberg, Germany) before the self-assembly process. In the case of XPS, XRR, and CD experiments, they were instead treated with UV-ozone (UV Ozone Cleaner UVC-1014, NanoBioAnalytics, optical power 4 W and 254 nm) for 15 minutes. Samples were immersed in a 10 mM solution of 1,1'-binaphthyl-2,2'-diyl hydrogenphosphate (BNP) solution in tetrahydrofuran (THF) or ethanol for 72 h. *R-, S-,* and *rac*-BNP were obtained from Sigma-Aldrich. After immersion, samples were annealed at 80 °C for 1 h, rinsed with isopropanol, and again annealed at 80 °C for 10 min. SAM-coated samples were fabricated for both chiral enantiomers (*R*-BNP and *S*-BNP) and a racemic mixture of both enantiomers (*rac*-BNP).

## 2. X-ray Photoelectron Spectroscopy (XPS)

The measurements were carried out at the bending magnet HE-SGM beamline of the synchrotron storage ring BESSY II in Berlin, Germany. This beamline provides a linearly polarized X-ray light with a polarization factor of ~90 %. A custom-designed experimental station was used.[1] All measurements were performed at room temperature. The experiments were conducted in ultra-high vacuum, at a base pressure of ca. $1\times10^{-9}$ mbar. The XPS spectra were measured with a Scienta R3000 electron energy analyzer, in normal emission geometry. The primary photon energy (PE) was set to 350 or 580 eV to access specific core levels and to vary the surface sensitivity. The energy resolution at these PEs was ~0.3 eV and ~0.5 eV, respectively. The binding energy (BE) scale of the spectra was referenced to the Au $4f_{7/2}$ emission of the Au substrate at 84.0 eV.[2] For the evaluation of the XPS data, we used the standard expressions for the attenuation and self-attenuation of the photoemission signals from the substrate and an overlayer[3] and the literature values for the attenuation lengths of these signals in SAM-like films.[4] The data for the binol SAMs were referenced to those for *n*-hexadecanethiolate (C16) monolayer on Au(111). This particular system serves as an established reference for SAM studies and is characterized by a thickness of 1.89 nm and a packing density of $4.63 \cdot 10^{14}$ molecules·cm$^{-2}$.[5, 6]

The thickness of the substrate oxide film, $d_{\mathrm{NiO}}$, was estimated based on the relative weights of the oxide and metal contributions in the Ni 3p spectra of the samples (see Figure S1 as an example).

$$d_{\mathrm{NiO}} = \lambda \ln[(I_{\mathrm{NiO}} + I_{\mathrm{Ni}})/I_{\mathrm{Ni}}], \quad (1)$$

where $\lambda$ is the attenuation length of the Ni 3p signal at the given kinetic energy (1.1 nm)[4] and $I_{\mathrm{NiO}}$ and $I_{\mathrm{Ni}}$ are the relative intensities of the Ni oxide and Ni metal signals, respectively. The derived $d_{\mathrm{NiO}}$ values varied from 1.07 to 1.47 nm for the different samples, with the average value of 1.3 nm.

The thicknesses of the BNP and C14 SAMs on Ni (hydrocarbon matrix only) were calculated based on the intensity of the C 1s signal for these films, referenced to the value for C16/Au. The packing densities of the BNP SAMs were calculated based on the intensities of the P 2p signals, referenced to the value for C14/Ni. The different attenuation of the P 2p signal by the binol and alkyl matrices was accounted for. It was assumed that the packing density of C14/Ni is similar to that of C16/Au.

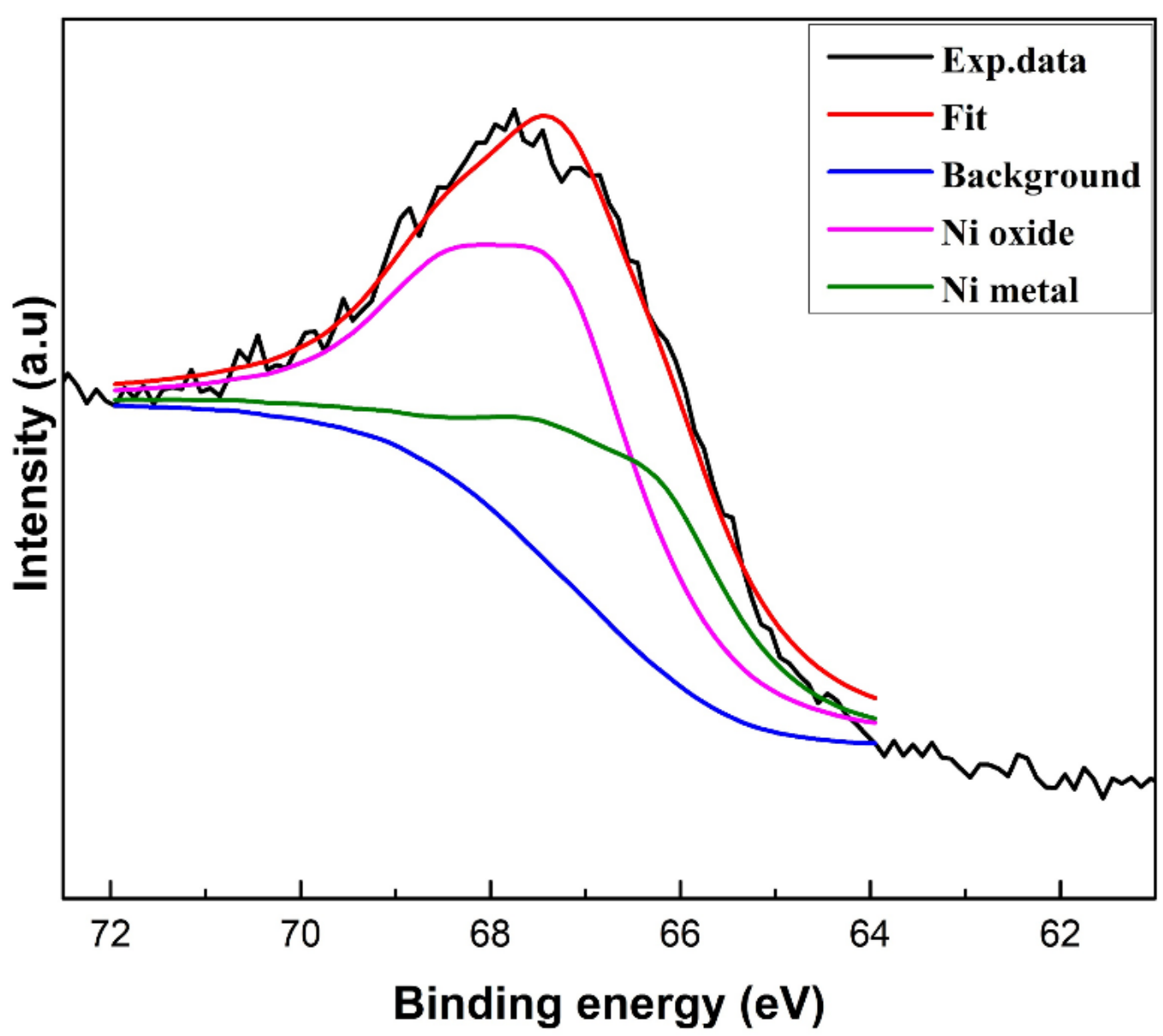

**Figure S1:** As measured Ni 3p XP spectrum of *R*-BNP on Ni, representative of the entire sample series. The spectrum is decomposed in the contributions associated with metal Ni and $NiO_x$.

3. Near-edge X-ray Absorption Fine Structure Spectroscopy (NEXAFS)

The NEXAFS spectra were collected at the carbon K-edge in the partial electron yield mode with a retarding voltage of −150 V. The incidence angle of the primary X-rays was varied between the normal (90°; the **E** vector parallel to the sample surface) and grazing (20°; the **E** vector nearly perpendicular to the sample surface) incidence geometry to monitor the linear dichroism reflecting the molecular orientation in the SAMs.[7] The energy resolution was ~0.3 eV. The photon energy (PE) scale was referenced to the pronounced π* resonance of HOPG at 285.38 eV.[8] The raw spectra were corrected for the PE dependence of the incident photon flux and reduced to the standard form with zero intensity in the pre-edge region and the unity jump in the far post-edge region.

Representative C K-edge NEXAFS spectra of *S*-BNP, *rac*-BNP, and C14 on Ni are presented in Figure S2. The spectrum of the reference C14/Ni in Fig. S2c shows the characteristic resonances of the alkyl chains[9–11], meaning C-H band at ~287.7 eV, comprised of several σ*C-H / Rydberg resonances, and σ* C-C and σ* C-C′ resonances at ~293.4 eV and ~302 eV, respectively. These resonances exhibit a pronounced linear dichroism characteristic of well-ordered alkyl chains in upright orientation.[11] Specifically, the C-H band, with the transition dipole moment (TDM) perpendicular to the alkyl backbone[12], has a higher intensity at **E** parallel to the sample surface (90°). The extent of the linear dichroism in the spectra of C14/Ni (Figure S2c) is similar to that for C16/Au (Figure S3), suggesting a similar molecular orientation with an average tilt angle of 30-35°, and supports the assumption about a similar packing density in these SAMs. A further important implication is the possibility of preparing SAMs of comparable quality by the phosphoric acid anchoring on oxidized Ni, as by thiolate anchoring on Au(111). Analogous quality can therefore be expected for BNP, which is indeed supported by the spectra *S*-BNP/Ni and *rac*-BNP/Ni in Figures S2a and S2b, respectively, showing the characteristic absorption features of naphthalene[13–15] – the major building block of BNP. It is above all the double $\pi_{1a}$* and $\pi_{1b}$* resonances at ~284.75 eV and ~285.7 eV, with the characteristic intensity relation between both peaks, followed by weaker and less pronounced π* and σ* resonances at ~288.5 eV, ~290.2 eV, ~293.8 eV, and ~301 eV. Significantly, the difference spectrum of *S*-BNP/Ni exhibits a small but pronounced linear dichroism at the positions of the $\pi_{1a}$* and $\pi_{1b}$* resonances, which indicates a certain degree of orientational order in this monolayer. Considering that the respective TDMs are oriented perpendicular to the plane of the naphthalene moieties, higher intensities of these resonances at **E** parallel to the sample surface (90°) suggest a predominant upright orientation in these films. A qualitative evaluation of the NEXAFS data following the formalism for a vector-like orbital[7, 11] gave an average tilt angle of the naphthalene moieties in the backbone of the *S*-BNP/Ni at ∼31 ± 3 °. In contrast, there is no dichroism in the spectra of the *rac*-BNP, indicating that this film is most likely disordered and of lower quality than the enantiomer monolayers, in full agreement with the XPS data.

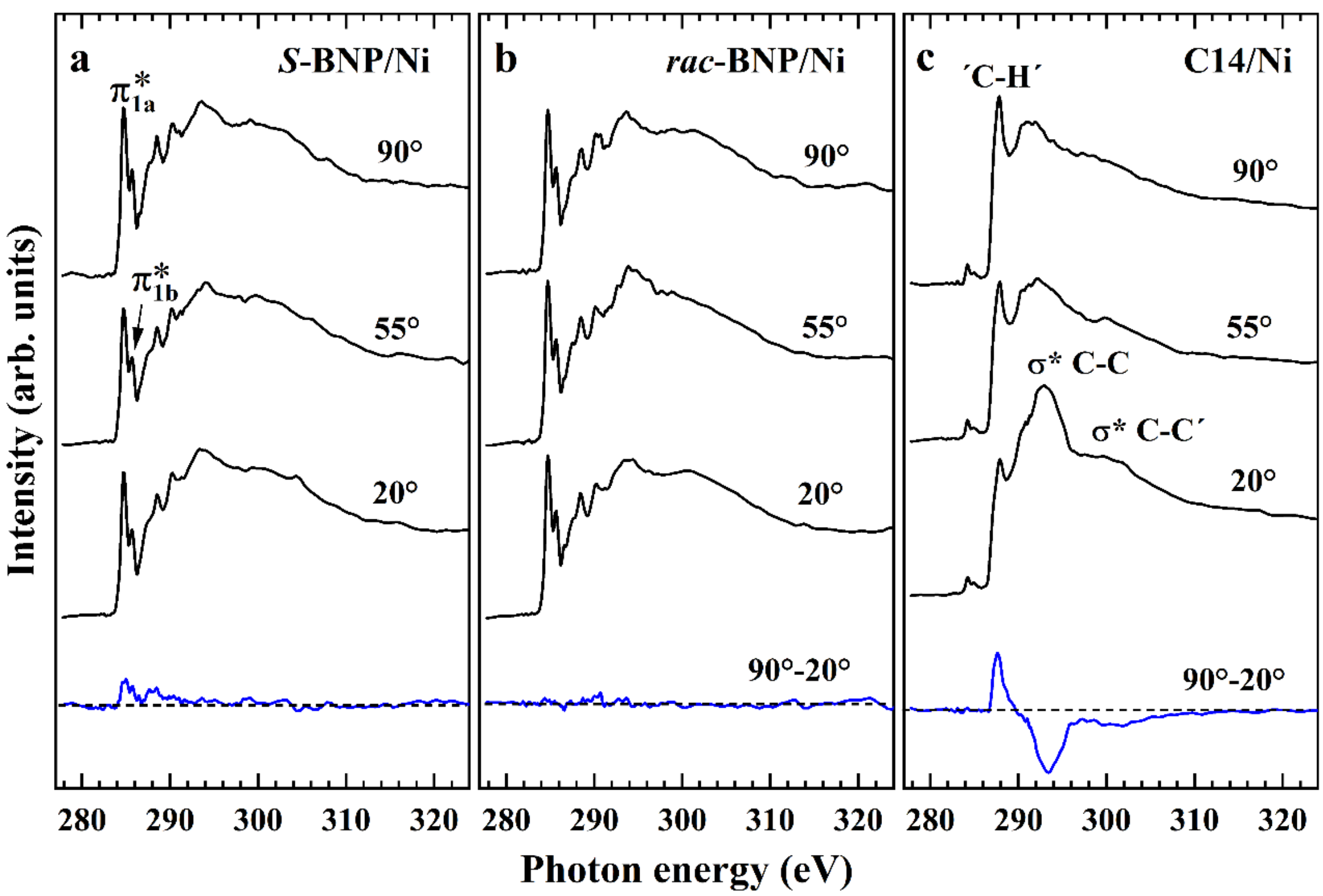

**Figure S2:** C K-edge NEXAFS spectra of *S*-BNP (a), *rac*-BNP (b), and C14 (c) on Ni (the spectra of *R*-BNP are very similar to that of *S*-BNP). The data include the spectra acquired at the different X-ray incidence angles, marked at the spectra (black lines), and the difference spectra resulting from subtracting the spectrum acquired at an X-ray incidence angle of 90° from that acquired at 20° (blue lines). The most prominent resonances are assigned. Horizontal, black dashed lines correspond to zero.

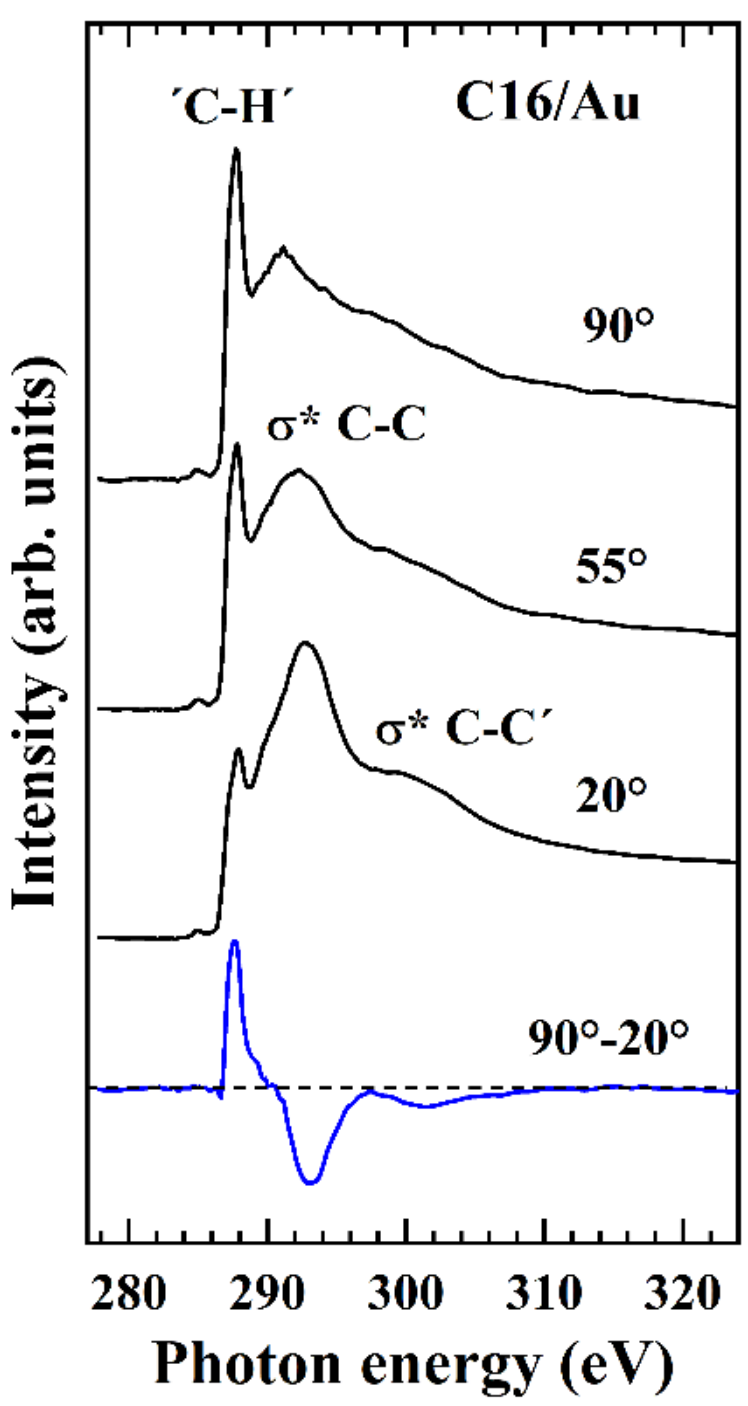

**Figure S3:** C K-edge NEXAFS spectra of C16/Au. The data include the spectra acquired at the different X-ray incidence angles, marked at the spectra (black lines), and the difference spectra resulting from subtracting the spectrum acquired at an X-ray incidence angle of 90° from that acquired at 20° (blue lines). The most prominent resonances are assigned. Horizontal, black dashed lines correspond to zero.

4. Atomic Force Microscopy

To characterize the surface morphology of the BNP SAMs on nickel, AFM in both tapping and contact mode was used. More specifically, the Dimension V AFM (Bruker/Veeco, Billerica, Massachusetts, United States) was utilized with the diamond-like-carbon-coated tips 190DLC (tip radius: 15 nm) from BudgetSensors (Sofia, Bulgaria). The instrument was controlled, and the images were recorded using the Nanoscope 7 software. In tapping mode, 5 × 5 $\mu m^2$ images were taken. To determine the SAM thickness, the so-called "scratching" method was applied on a 1 × 1 $\mu m^2$ area.[16] The voltage setpoint that applied the force to achieve this was calibrated so that the bare nickel substrate below was not damaged. Subsequently, a 3 × 3 $\mu m^2$ tapping image was taken. The height difference between the "scratched" and "unscratched" areas of the image indicates the SAM thickness. The AFM data was analyzed using Nanoscope and Origin.

Tapping measurements of both bare and SAM-coated samples are shown in Figure S4. The bare Ni surfaces (Fig. S4a) exhibit an average root-mean-square (RMS) roughness of $r_{\mathrm{RMS}} = 0.6 \pm$ 0.1 nm. For the three SAM-coated samples (*R*-BNP in Fig. S4b, *S*-BNP in Fig. S4c, and *rac*-BNP in Fig. S4d), some agglomerations are visible, indicating that the molecules did not form a perfect monolayer in these regions. Because of these spots, the average RMS roughness increases (e.g., $r_{RMS} > 1$ nm for the *R*-BNP sample in Fig. S4b). However, in the areas in between these presumed molecule agglomerations, the roughness is in the same range as for the bare Ni surface (Fig. S4a). Therefore, no difference in roughness can be observed for bare Ni surfaces compared to surface areas with expected SAM. Given the low molecular length of BNP (≈ 1 nm), this finding aligns with previous publications and indicates the presence of well-ordered SAMs: Typically for phosphonic acids, the roughness of TiN substrates was preserved after the deposition of a roughly 0.8 nm thick SAM of 6-aminohexylphosphonic acid. Further, for Al substrates of much higher roughnesses (> 30 nm), no significant change in roughness was reported after the growth of alkyl phosphonic acid monolayers.[17] Similarly, for thiol-based SAMs, it was shown that neither the much longer alkane chain 16-mercaptohexadecanoic acid nor the thiolated biphenyl derivative 1, 1'-biphenyl-4-thiol (BPT) altered the roughness of an Au substrate.[18] A monolayer of the cross-linked version of BPT even led to a decrease in roughness.[18, 19]

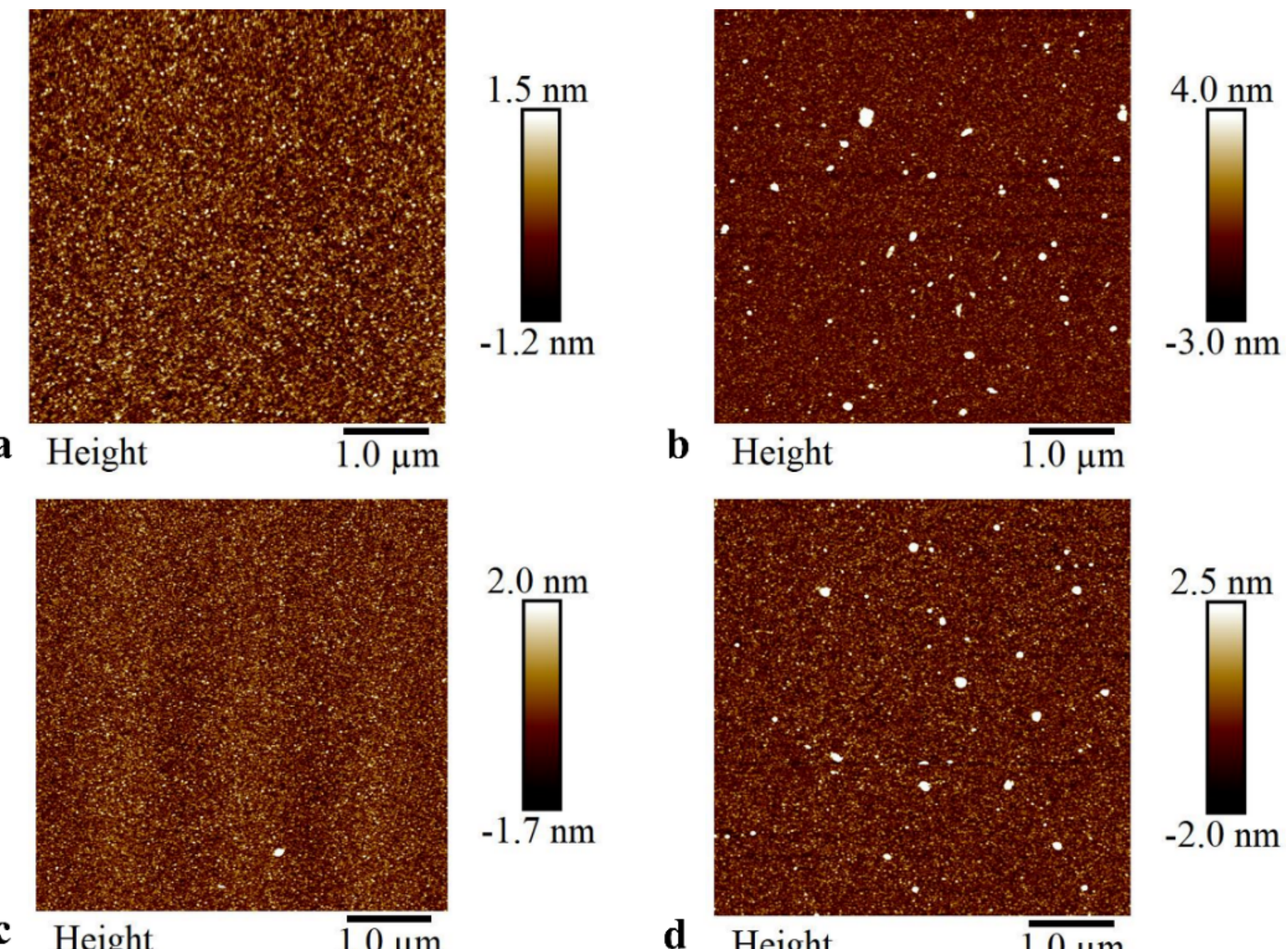

**Figure S4:** Representative tapping mode images for the bare Ni substrate (a) and Ni substrates coated with *R*-BNP (b), *S*-BNP (c), and the racemic mixture *rac*-BNP (d).

The results of the "scratching" experiments are presented in Figure S5. The recordings Fig. S5a and Fig. S5b depict the tapping-mode images after scratching *R*-BNP and *rac*-BNP-coated samples, respectively. As is visible, a central square of lower height is apparent for all three SAMs (see Fig. 3d, main paper), indicating successful mechanical removal of the soft organic layer. Height variations along the horizontal axes of the images are averaged over the scratched areas and plotted in Fig. S5c and Fig. S5d for *R*-BNP and *rac*-BNP, respectively. Here, areas with agglomerations left in the supposedly scratched areas are omitted.

The resulting SAM thickness determinations – as calculated via the height differences between the scratched and unscratched parts in the profiles – are presented in Table S1. The height peaks to the left and right of the scratched center are caused by the accumulation of molecules being pushed aside by the AFM tip and are, therefore, not considered for the calculation. As can be seen, all calculations via the height profile yield a layer thickness ≥ 0.5 nm. The scratching experiment for the *R*-BNP samples (Fig. S5d) results in a thickness of ~ 0.8 nm, which matches closest with the theoretical molecule length of 0.96 nm.

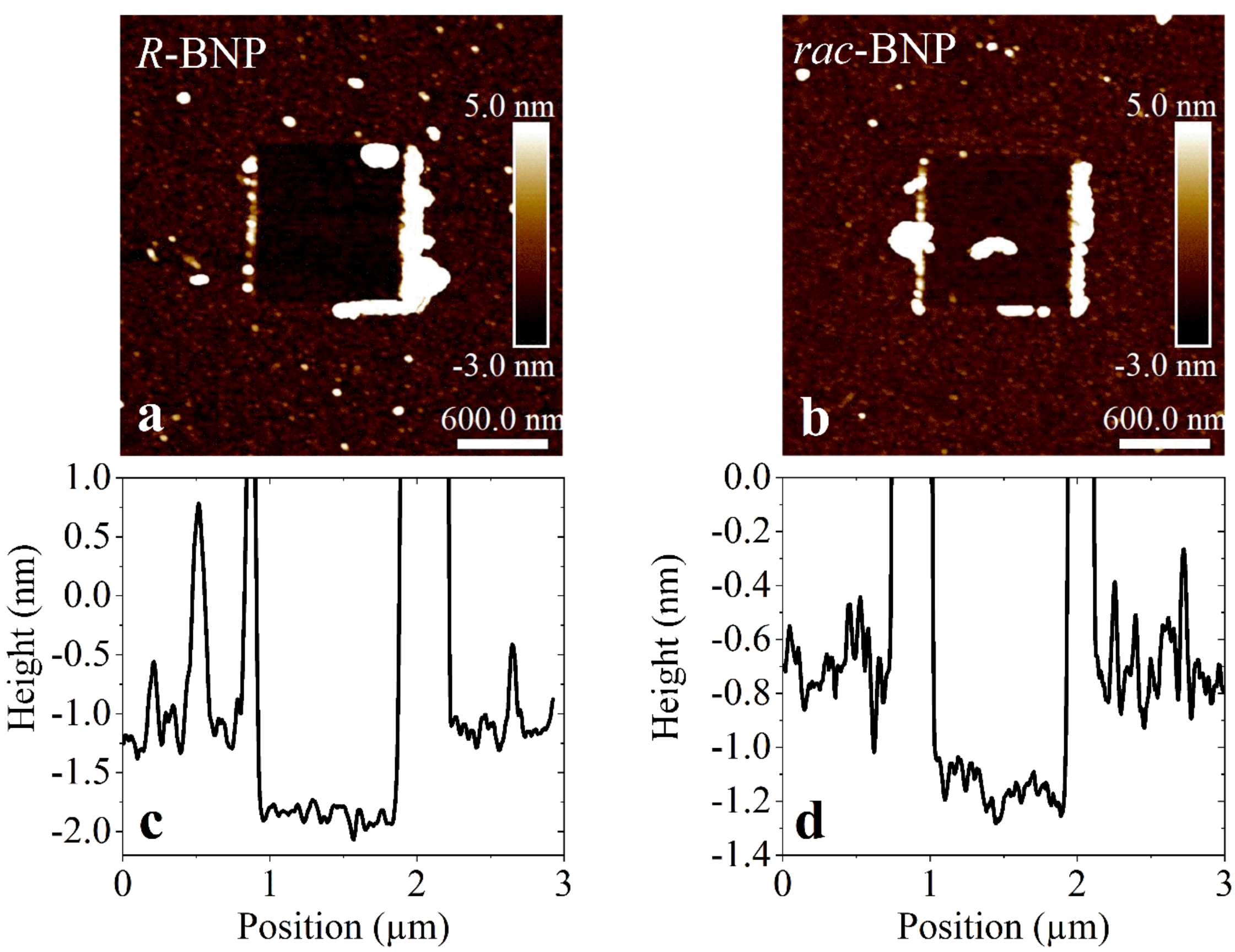

**Figure S5:** AFM scratching experiments. Upper row: (a) and (b) depict the 3 × 3 µm$^2$ tapping recordings of scratched samples of *R*-BNP and *rac*-BNP, respectively. Lower row: mean height profiles along the horizontal axis over the scratched areas in (a) and (b) are depicted in (c) (*R*-BNP) and (d) (*rac*-BNP), respectively. Here, horizontal profiles that include molecule agglomerations in the central scratched area (as in the center in (b)) were omitted.

**Table S1:** SAM thickness estimated from height profile (Fig. S5c, Fig 3e, and Fig. S5d). Error represents the largest standard deviation of the height profile of the respective scratched or unscratched areas.

| SAM molecule | SAM thickness [nm] determined via height profile |
|---|---|
| *R*-BNP | 0.8 +/- 0.5 |
| *S*-BNP | 0.6 +/- 0.1 |
| *rac*-BNP | 0.5 +/- 0.1 |

**5. X-ray Reflectivity**

The X-ray reflectivity measurements were performed using a SmartLab diffractometer (Rigaku Corporation, Tokyo, Japan), which utilized monochromatic Cu $K\alpha_1$ radiation ($\lambda = 1.54056$ Å). The setup includes a divergent slit with parallel-beam optics and a 2 mm Soller slit. Specular $\theta$-$2\theta$ scans were collected over a $2\theta$ range of 0° to 10°, with a step size of 0.01° and a counting time of 0.5 deg·min$^{-1}$. The resulting reflectivity profiles were analyzed and fitted using the GenX software package.

**Table S2:** Estimated thicknesses, roughness, and densities of the layers based on the XRR data.

| Layer | Thickness (nm) (±0.06) | Roughness (nm) (±0.01) | Density (g/cm$^3$) (±0.532) |
|---|---|---|---|
| $SiO_2$ | 2.48 | 0.29 | 2.39 |
| Ti | 10.04 | 0.50 | 4.77 |
| Ni | 99.19 | 1.68 | 8.77 |
| NiO | 1.33 | 0.76 | 4.71 |
| BNP | 0.98 | 0.63 | 1.73 |

**Table S3:** Comparison of BNP monolayer thickness values obtained from XRR, AFM scratching, and XPS measurements.

| | **Theoretical** (molecule contour length) | **XPS** | **XRR** (*rac*-BNP) | **AFM Scratching** (*R*-, *S*-, *rac*-BNP) |
|---|---|---|---|---|
| BNP SAM thickness (nm) | 0.96 | 0.92 | 0.98 ± 0.09 | 0.4 – 0.8 ± 0.5 |

The surface density of *R*- and *S*-BNP of $\sim 1.9 \cdot 10^{14}$ molecules·cm$^{-2}$ on NiO/Ni substrates obtained from the XPS measurements agrees very well with the crystal structures:[20, 21] enantiopure *R*- and *S*-BNP crystallize in a layered structure (space group orthorhombic $P\,2_12_12_1$) with the polar phosphoric acid moieties separated from the aromatic binaphthyl part of the molecules. These layers have orthogonal lattice constants of 0.919 nm and 0.599 nm, indicating a packing density of 1.817 molecules·nm$^{-2}$ within the layers.

In contrast, racemic BNP (space group triclinic *P*-1)[22] does not crystallize in layers like the enantiopure compounds. Therefore, the crystal structure cannot be used as a reference for surface-adsorbed *rac*-BNP. This difference between the structures of enantiopure and racemic BNP might account for the lower experimental packing density of *rac*-BNP as determined via XPS.

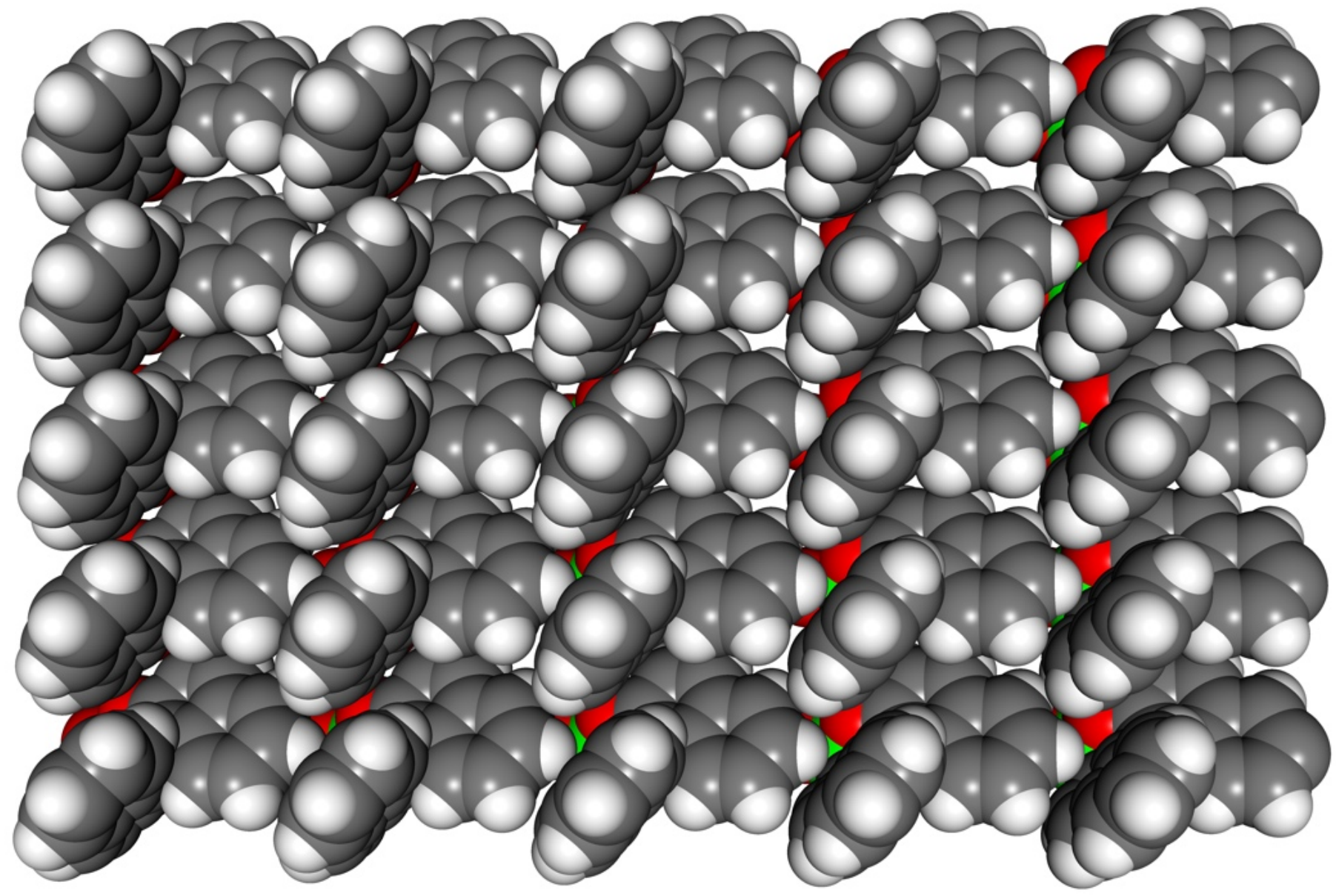

**Figure S6:** Top view onto a layer of *S*-BNP (4.6 x 3.0 $nm^2$), sliced out of the crystal structure,[21] with a packing density of 1.817 molecules·$nm^{-2}$. A similar structure can be assumed for the monolayer adsorbed to nickel oxide and other substrates.

## 6. Circular Dichroism in Solution

Spectra were recorded with the same instrument as for thin films (see below). The measurements were conducted in solution at room temperature, utilizing a quartz cuvette with a path length of 1 cm and a sample concentration of 0.01 mM in ethanol. Spectra were collected over a wavelength range of 205 to 500 nm, with a scan speed of 1000 nm·$min^{-1}$, a bandwidth of 1 nm, a digital integration time (D.I.T.) of 0.125 seconds, and a data pitch of 0.5 nm. Each spectrum was obtained by averaging five accumulations to enhance the signal-to-noise ratio. The solvent baseline was recorded separately and then subtracted from the sample spectra. Absorbance values were converted to molar extinction coefficients (ε, $M^{-1}$·$cm^{-1}$) using the Beer-Lambert law. CD ellipticity ($\theta,$ mdeg) was converted to molar circular dichroism (Δε, $M^{-1}$·$cm^{-1}$) using the relation: $\Delta\varepsilon = \theta/(32980 \cdot c \cdot l)$, where $c$ in the concentration and $l$ is the path length.

The UV-vis absorption spectra for *R*-BNP, *S*-BNP, and *rac*-BNP are presented in Figure S7 (taken with same instrument as CD). All samples show strong absorption that extends to approximately 350 nm, primarily attributed to intense π-π* transitions of the binaphthyl moiety in the deep UV region (less than 240 nm).[23–25] A strong absorption maximum occurs at 215.5 nm with an absorbance $A$ of approximately 1.20, which corresponds to a molar extinction coefficient $\varepsilon = 1.2 \cdot 10^5$ $mol^{-1}$·$cm^{-1}$. There is also a weaker shoulder at approximately 228 nm, with an absorbance of around 0.90, resulting in an extinction coefficient of $\varepsilon = 9.0 \cdot 10^4$ mol$^{-1}$·cm$^{-1}$. Beyond 320 nm, the spectra decline rapidly, with absorbance values of $A \leq 0.10$ and $\varepsilon \leq 1.0 \cdot 10^4$ mol$^{-1}$·cm$^{-1}$. The absorption profiles of the racemic and enantiopure BNP samples are nearly identical, indicating that chirality does not affect the ground-state electronic transitions of BNP.

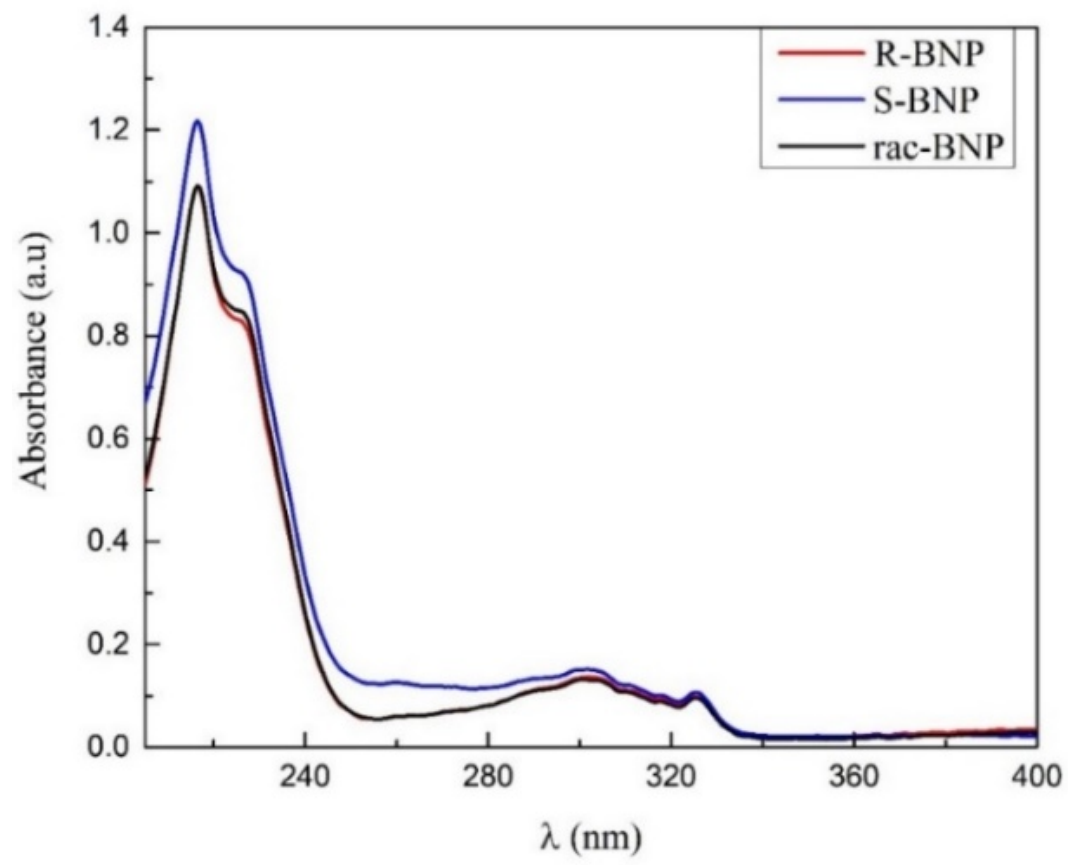

**Figure S7:** UV absorption spectra of *R*-BNP (red), *S*-BNP (blue), and *rac*-BNP (black) in 0.01 mM ethanol solution

**7. Circular Dichroism (CD) of Thin Films**

For the substrate, 30 nm Ni was sputtered onto sapphire samples (C-cut(001), 0.5 mm, double side polished, supplied by CrysTec) using a home-built radio frequency sputtering system. Ar was used as the working gas. BNP was deposited on the Ni films using the above-described SAM deposition procedure. CD thin film measurements were performed using a JASCO J-1500 CD spectrophotometer (JASCO Corporation, Tokyo, Japan). Spectra were collected in the wavelength range of 205-500 nm with a scan speed of 200 nm·min$^{-1}$, a data pitch of 0.1 nm, a digital integration time of 2 s, and a bandwidth of 1 nm. Samples were mounted on a custom-built rotation stage that enabled automated rotation in 15° steps between azimuthal angles of 0° and 180°. Measurements were acquired for both front and back configurations. To extract the intrinsic CD response, spectra are averaged over all rotation angles as well as the front and back measurements. Furthermore, the spectrum of a bare Ni reference sample was subtracted, isolating the contribution of the BNP layer.

For comparison, the UV-Vis and CD spectra for *S*-BNP were also calculated by time-dependent density functional theory (TD-DFT) on the TD-CAM/6-311+G(d,p) level of theory.[26] The calculations provided the same signs for the respective CD signals, thus confirming the absolute configuration of the compound.

## 8. Magnetic Conductive Atomic Force Microscopy

### 8.1. Method Description

The chiral-induced spin selectivity effect was measured via its induced magnetoresistance (CISS-MR) using a custom-built mc-AFM setup. The system includes a Beetle Ambient AFM and an electromagnet controlled by an R9 electronic controller (RHK Technology, Troy, MI, USA). Pt-coated conductive tips of the type DPE-XSC11 (MikroMasch, Sofia, Bulgaria) with a spring constant of 3 - 5.6 N $m^{-1}$ and a tip radius < 40 nm were utilized. The presented current-voltage (I-V) characteristics are based on 40-62 I-V sweeps for both magnetic field directions, while the tip was kept over the same position. During the measurements, a magnetic field of ~ 0.5 T was applied upwards or downwards perpendicular to the sample plane. The response of the magnetic film to that applied field can be seen in the SQUID measurements (Figure S13 and S14). The applied force to bring the tip in contact with the surface was set to 8-10 nN and kept constant (on a given spot) for both magnetic field directions.

### 8.2. Statistics

The IV characteristics presented in the main paper are based on five measurement positions for *S*-BNP (individual measurements shown in Figure S8), three measurement positions for *R*-BNP (individual measurements shown in Figure S9), and four positions for *rac*-BNP (individual measurements shown in Figure S10).

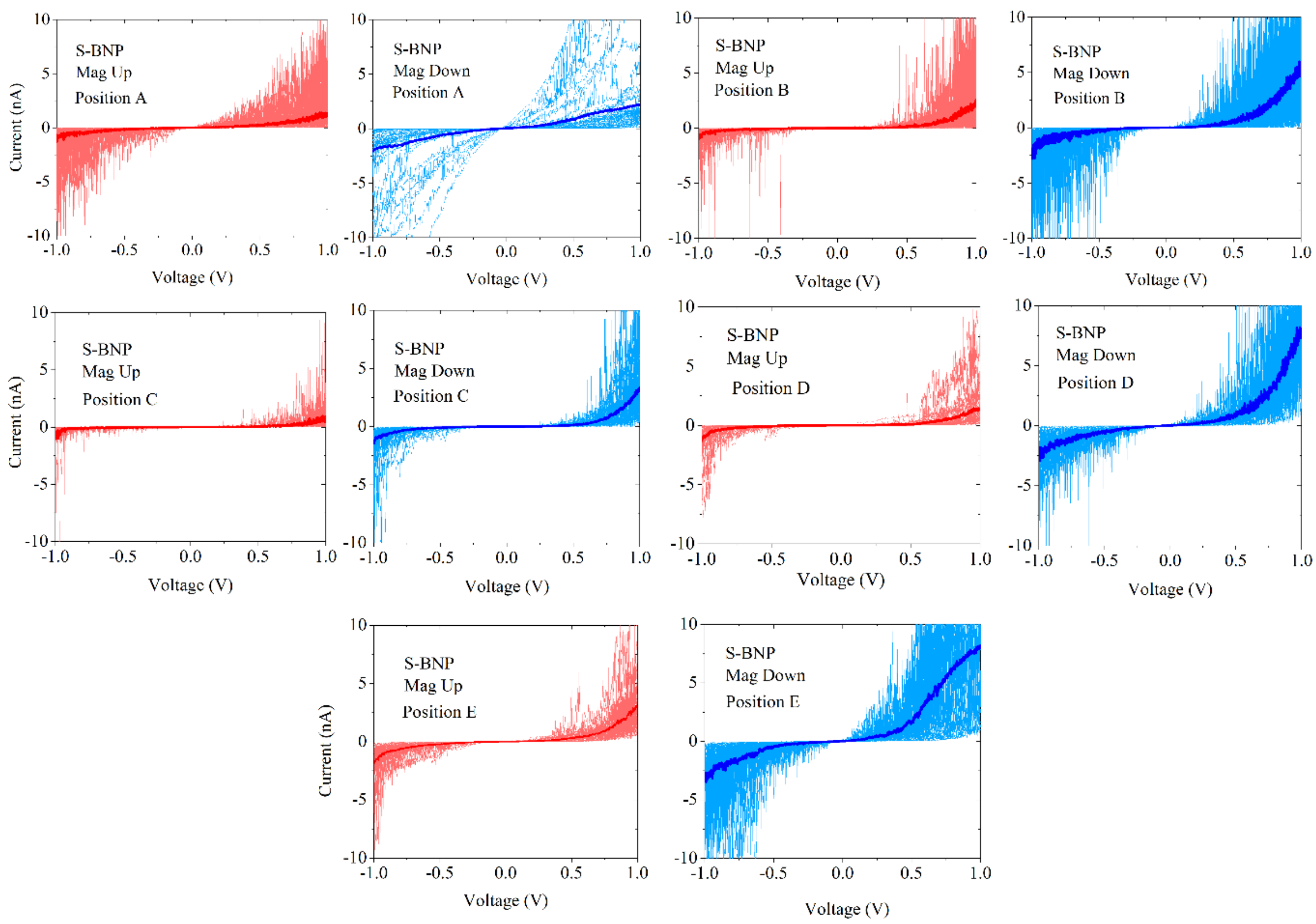

**Figure S8:** Individual current-voltage measurements for *S*-BNP-coated sample. Red (blue) lines indicate individual curves taken while the magnetic field was pointing upwards (downwards) perpendicular to the sample plane. Thicker lines indicate the respective mean current for that position. The following number of curves were taken for upwards (downwards) magnetization: Position A: 61 (55), Position B: 55 (54), Position C: 22 (66), Position D: 55 (28), Position E: 60 (63).

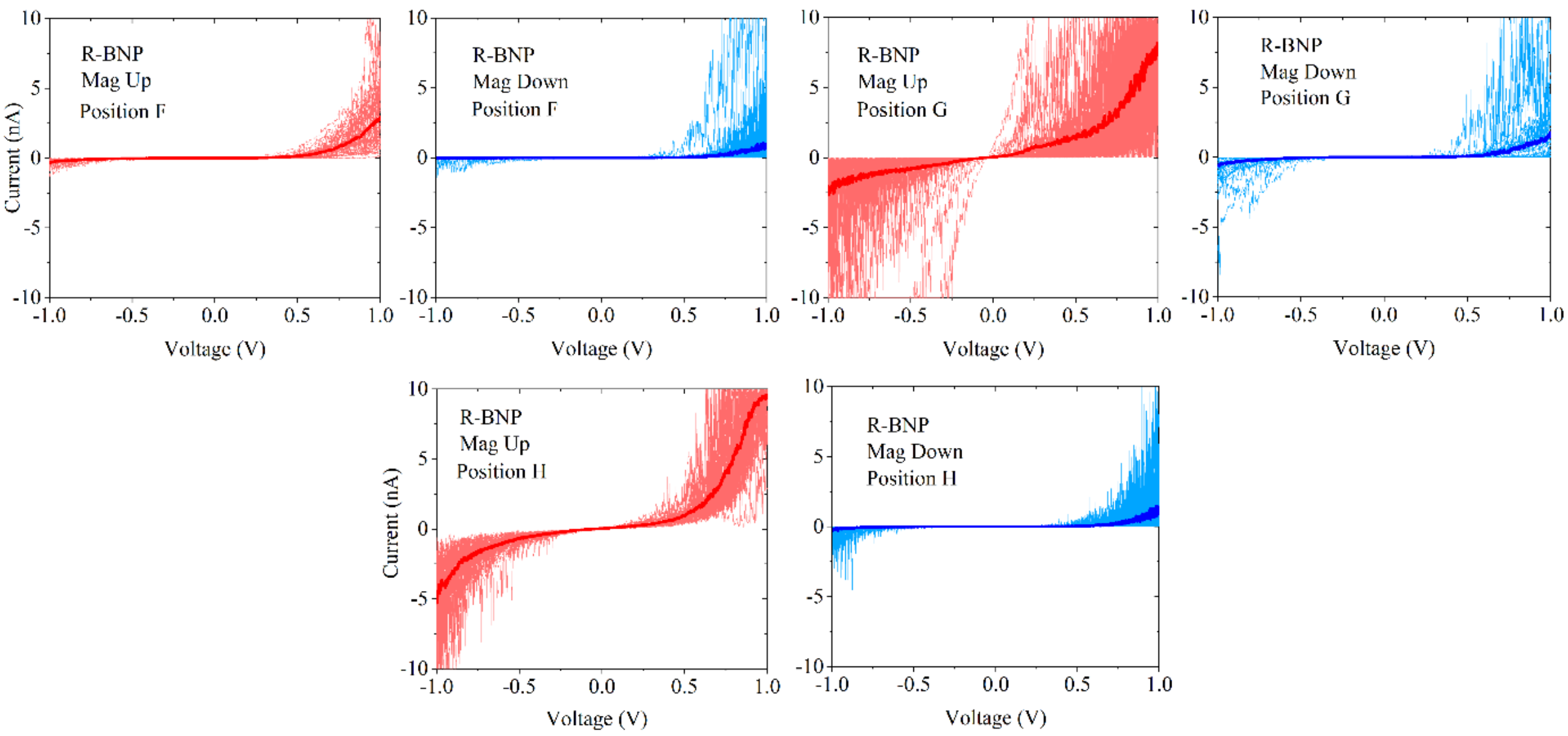

**Figure S9:** Individual current-voltage measurements for *R*-BNP-coated sample. Red (blue) lines indicate individual curves taken while the magnetic field was pointing upwards (downwards) perpendicular to the sample plane. Thicker lines indicate the respective mean current for that position. The following number of curves were taken for upwards (downwards) magnetization: Position F: 62 (54), Position G: 50 (54), Position H: 51 (63)

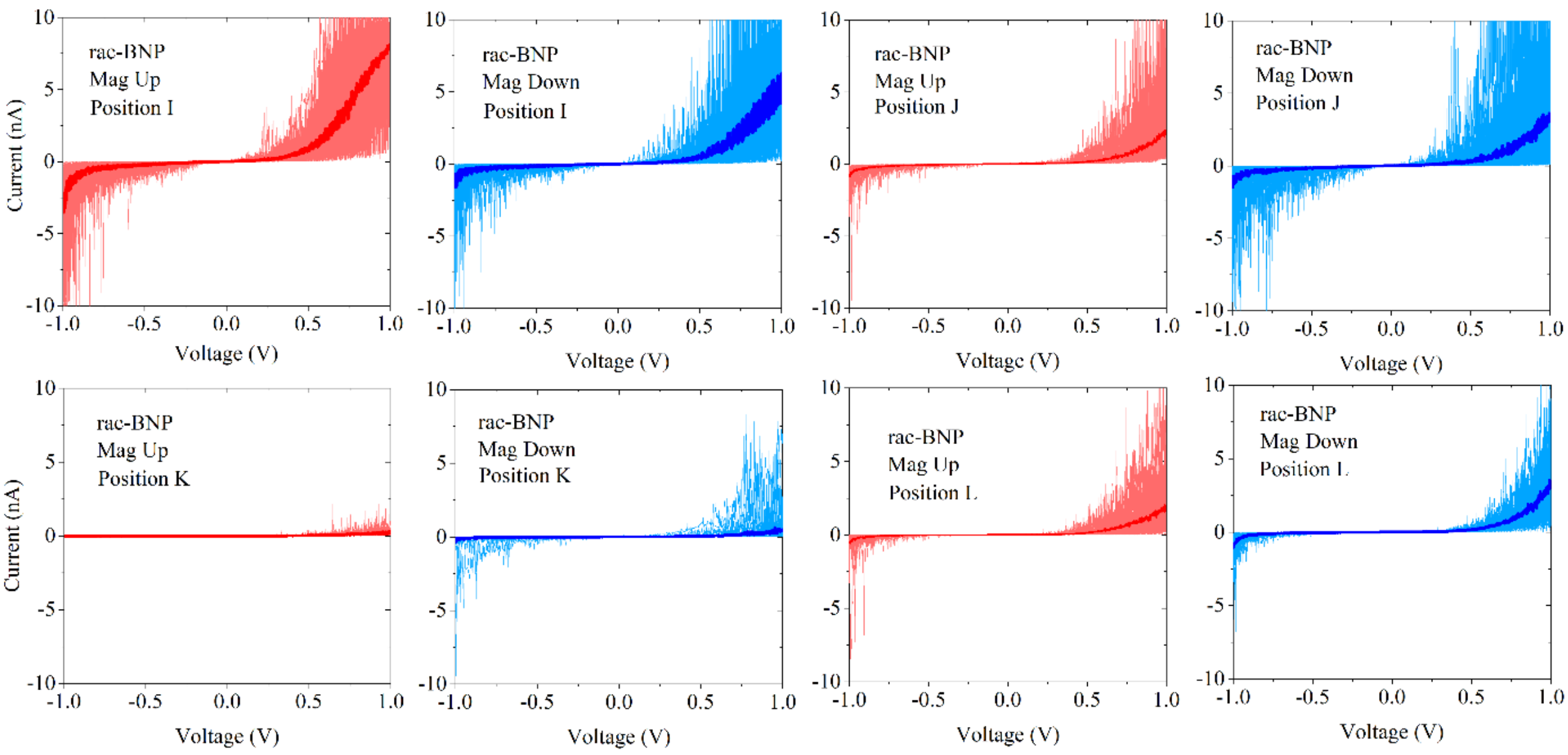

**Figure S10:** Individual current-voltage measurements for *rac*-BNP-coated sample. Red (blue) lines indicate individual curves taken while the magnetic field was pointing upwards (downwards) perpendicular to the sample plane. Thicker lines indicate the respective mean current for that position. The following number of curves were taken for upwards (downwards) magnetization: Position I: 51 (54), Position J: 54 (54), Position K: 40 (51), Position L: 54 (52)

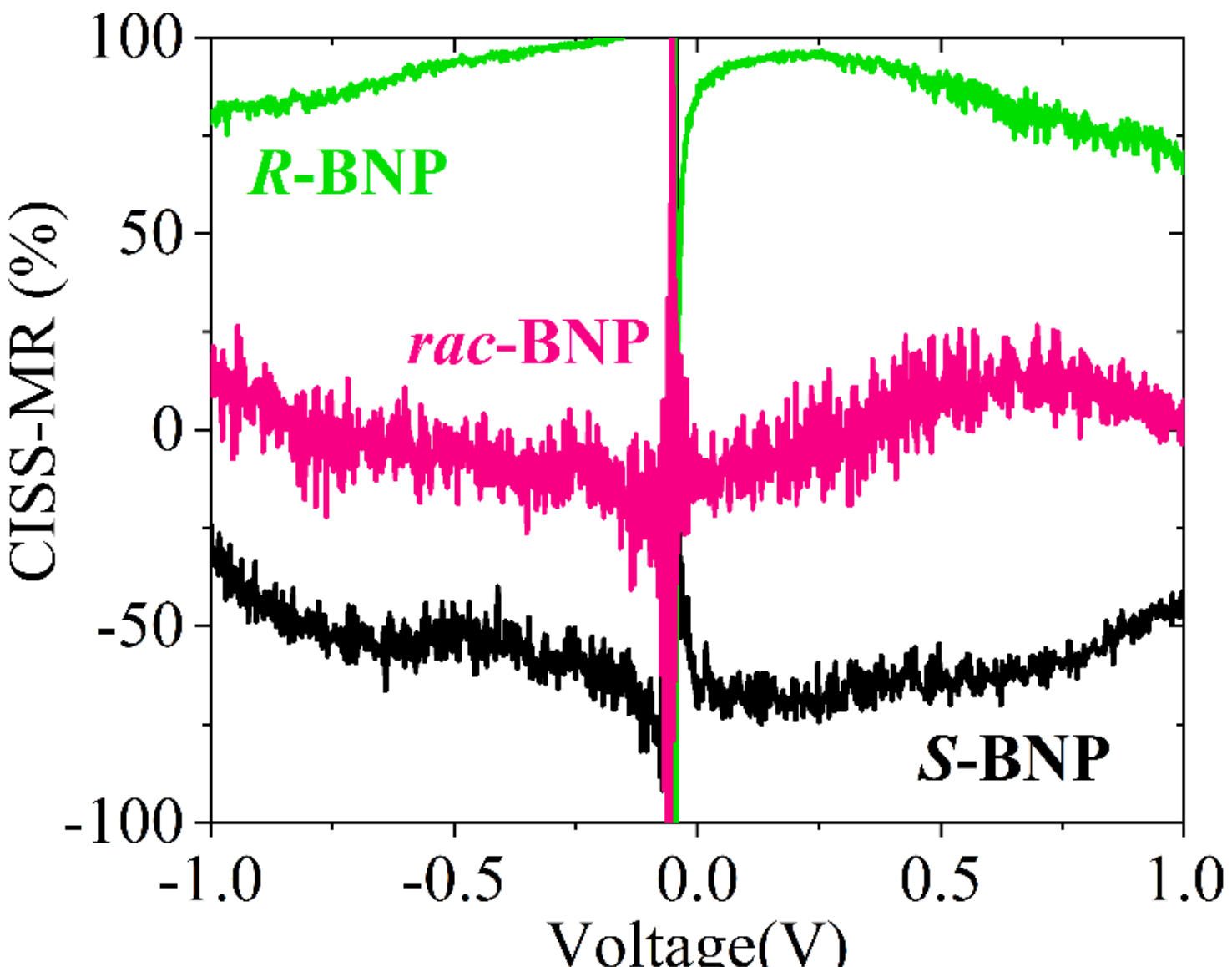

**Figure S11:** Calculated CISS-MR for *R*-BNP (green), *S*-BNP (black), and *rac*-BNP (magenta) over the entire voltage range -1 V to 1 V. A peak outside the shown y-axis is reached between -0.25 V and 0 V for both enantiomers and the racemic mixture. This is a numerical artifact, as both $I_{\text{Mag Up}}$ and $I_{\text{Mag Down}}$ reach 0 A. This peak not being exactly at 0 V could be a result of the molecule dipole or a slight offset in the mc-AFM measurement setup.

9. Fowler-Nordheim Tunneling Model

For the FN plot, only the data with positive voltages $V$ and currents $I$ are analyzed and transformed to produce a $\ln(I/V^2)$ vs $1/V$ plot. To find a linear fit for the FN regime, we follow the relation[27]

$$J = \alpha E^2 \cdot \exp\left(-\frac{B}{E}\right) \tag{2}$$

with

$$\alpha = \frac{m_e}{m^*} \frac{q^3}{8\pi h \phi_{\text{B}}} \tag{3}$$

and

$$B = \frac{8\pi}{3}\left(2\frac{m^*}{h^2}\right)^{1/2} \frac{\phi_{\text{B}}^{3/2}}{q}. \tag{4}$$

Here, $E$ is the electrical field, $J$ the current density, $\phi_{\text{B}}$ the effective barrier height, $q$ the electronic charge, $m_e$ the electron mass, $m^*$ the effective electron mass in the barrier material, and $h$ the Planck constant.[27] Rearranging equation (2) yields

$$\ln\left(\frac{I}{V^2}\right) = -Bd \cdot \frac{1}{V} + \ln\left(\frac{S\alpha}{d^2}\right). \tag{5}$$

For relation (5), we set $J = I/S$, where S is the area of the junction, and $E = V/d$, where $V$ is the applied voltage and $d$ is the barrier width (SAM and oxide). Consequently, the FN regime is fitted according to the linear equation $y = s \cdot x + c$, where $y = \ln(I/V^2)$, $s = -Bd$, $x = 1/V$, and $c = \ln(S\alpha/d^2)$. In a Python script, an ordinary least squares method is used to find $\theta = \begin{bmatrix} m \\ c \end{bmatrix}$, such that

$$\min_{\theta} \|A\theta - Y\|_2^2 , \tag{6}$$

where

$$Y = \begin{bmatrix} \ln(I_1/V_1^2) \\ \ln(I_2/V_2^2) \\ \vdots \\ \ln(I_N/V_N^2) \end{bmatrix}, \qquad A = \begin{bmatrix} 1/V_1 & 1 \\ 1/V_2 & 1 \\ \vdots & \vdots \\ 1/V_N & 1 \end{bmatrix}, \tag{7}$$

and $I_i$, $V_i$ are the datapoints in the FN regime for current and voltage, respectively.
To get the relative difference between the two effective barriers, we can use

$$\frac{\phi_{\mathrm{B}\uparrow}}{\phi_{\mathrm{B}\downarrow}} = \left(\frac{s_\uparrow}{s_\downarrow}\right)^{2/3} , \tag{8}$$

where $s_\uparrow$ and $s_\downarrow$ are the slopes of the fitted curves for the cases when the substrate is magnetized upwards and downwards, respectively. As can be seen, the resulting ratio of effective barriers is solely dependent on the two slopes and not on physical parameters. However, to derive the difference in the effective barrier height, a solution needs to be found for

$$|\Delta\phi_B| = |\phi_{B\uparrow} - \phi_{B\downarrow}|, \tag{9}$$

where $\phi_{B\uparrow}$ is the effective tunneling barrier for current measured when the ferromagnetic substrate is magnetized upwards and $\phi_{B\downarrow}$ the effective tunneling barrier for current measured when the ferromagnetic substrate is magnetized downwards. In combination with equation (8), this yields

$$|\Delta\phi_B| = \left|\phi_{B\downarrow}\left[\left(\frac{s_\uparrow}{s_\downarrow}\right)^{2/3} - 1\right]\right|, \tag{10}$$

Consequently, to get the effective barrier height difference, the full calculation results in

$$|\Delta\phi_B| = \left|\left|\frac{3qs_\downarrow}{8\pi d}\right|^{2/3} \cdot \left(\frac{h^2}{2m^*}\right)^{1/3} \cdot \left[\left|\frac{s_\uparrow}{s_\downarrow}\right|^{2/3} - 1\right]\right|. \tag{11}$$

For the approximation, the constants were chosen as $q = 1.602176634 \cdot 10^{-19}$ C, $h = 6.62607015 \cdot 10^{-34}$ J, and $m_e = 9.1093837015 \cdot 10^{-31}$ kg. For the remaining parameters, assumptions must be made. The tunneling barrier of the junction consists of two parts: the $NiO_x$ (~ 1.3 nm) and the BNP SAM (~ 1 nm). However, we will just model it as a single $d =$ 2.3 nm tunnel barrier with a uniform electric field over the whole junction. In previous literature, the electron mass in alkane SAMs was found to be $0.37m_e - 0.46m_e$ in a Simmons tunneling model.[28] For $SiO_2$ tunneling barriers, $0.5m_e$ shows to provide the best FN fit for the experimental data.[29] For the fitting in this work, the effective electron mass in the tunneling barrier is chosen to be $m^* \approx 0.5m_e$. The resulting FN plots with the associated fits are shown in Figure S12. For every FN plot, the linear fit was calculated based on the window $[1\ \mathrm{V}^{-1}, 2\ \mathrm{V}^{-1}]$. The derived fit parameters and the relations between $\phi_{B\uparrow}$ and $\phi_{B\downarrow}$ are shown in Table S4.

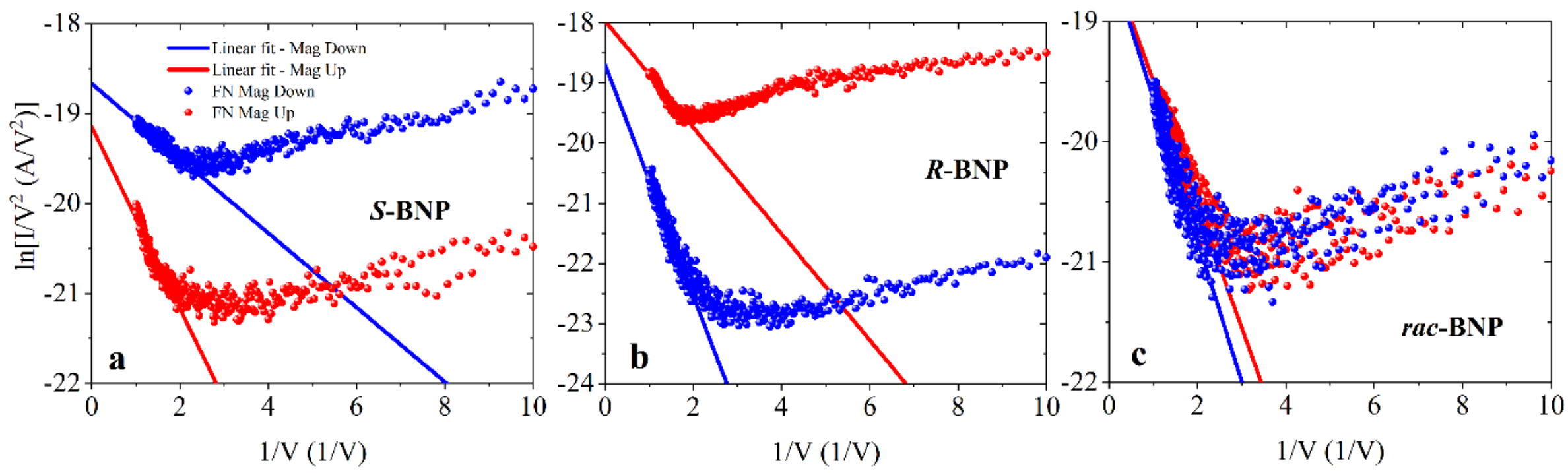

**Figure S12:** FN plot for the mc-AFM curves for *S*-BNP (a), *R*-BNP (b), and *rac*-BNP (c). Scatter plots indicate data points. Solid lines are the linear fits in the region [1 $V^{-1}$, 2 $V^{-1}$]. Blue color refers to the mean current measured when the substrate was magnetized downwards relative to the sample plane, red color refers to the mean current measured when the substrate was magnetized upwards relative to the sample plane.

**Table S4:** Fitting results for the FN linear fit: slope for linear fit for "Mag Up" ("Mag Down") current $s_\uparrow$ ($s_\downarrow$), offset for linear fit for "Mag Up" ("Mag Down") current $c_\uparrow$ ($c_\downarrow$), the ratio between the respective effective tunneling barriers ($\phi_{B\uparrow}/\phi_{B\downarrow}$), and the absolute difference between the two effective tunneling barriers $|\Delta\Phi_B|$.

| | $s_\uparrow$ [V] | $c_\uparrow$ | $s_\downarrow$[V] | $c_\downarrow$ | $\phi_{B\uparrow}/\phi_{B\downarrow}$ | $\lvert\Delta\phi_B\rvert$ |
|---|---|---|---|---|---|---|
| *R*-BNP | -0.88 | -18.0 | -1.92 | -18.7 | 0.6 | 126 meV |
| *S*-BNP | -1.01 | -19.2 | -0.41 | -18.7 | 1.8 | 91 meV |
| *rac*-BNP | -1.02 | -18.5 | -1.2 | -18.5 | 0.9 | 20 meV |

## 10. Superconducting quantum interference device magnetometry

The magnetic properties of the films were examined using a superconducting quantum interference device (SQUID) magnetometer (MPMS-XL, Quantum Design, San Diego, USA). Measurements were conducted at a temperature of 300 K, with the external magnetic field applied both out-of-plane (OOP) and in-plane (IP) relative to the substrate surface, reaching field strengths of up to 6 T. We subtracted the diamagnetic contributions from the substrate and sample holder from the raw data to ensure accuracy.

Figure S13 displays the magnetic hysteresis loops of a 100 nm Ni thin film, measured in both out-of-plane (OOP, Ni_OP) and in-plane (IP, Ni_IP) orientations. The normalized magnetization $M/M_S$ is plotted against the applied magnetic field in $\mu_0 H$. In both orientations, clear ferromagnetic hysteresis loops are observed, with magnetization saturating at approximately $\pm$ 800 mT for the OOP configuration and $\pm$ 400 mT for the IP configuration. This means that with the field applied during mc-AFM measurements, the substrate is not fully saturated, which could lead to an underestimation of the CISS-MR. These values are consistent with previously reported magnetization data for Ni thin films.[30–32] The coercive fields were determined to be 12.2 mT for the OOP orientation and 21.1 mT for the IP orientation, confirming the presence of well-defined ferromagnetic ordering. The distinct differences between the IP and OOP loops highlight the magnetic anisotropy of the film.

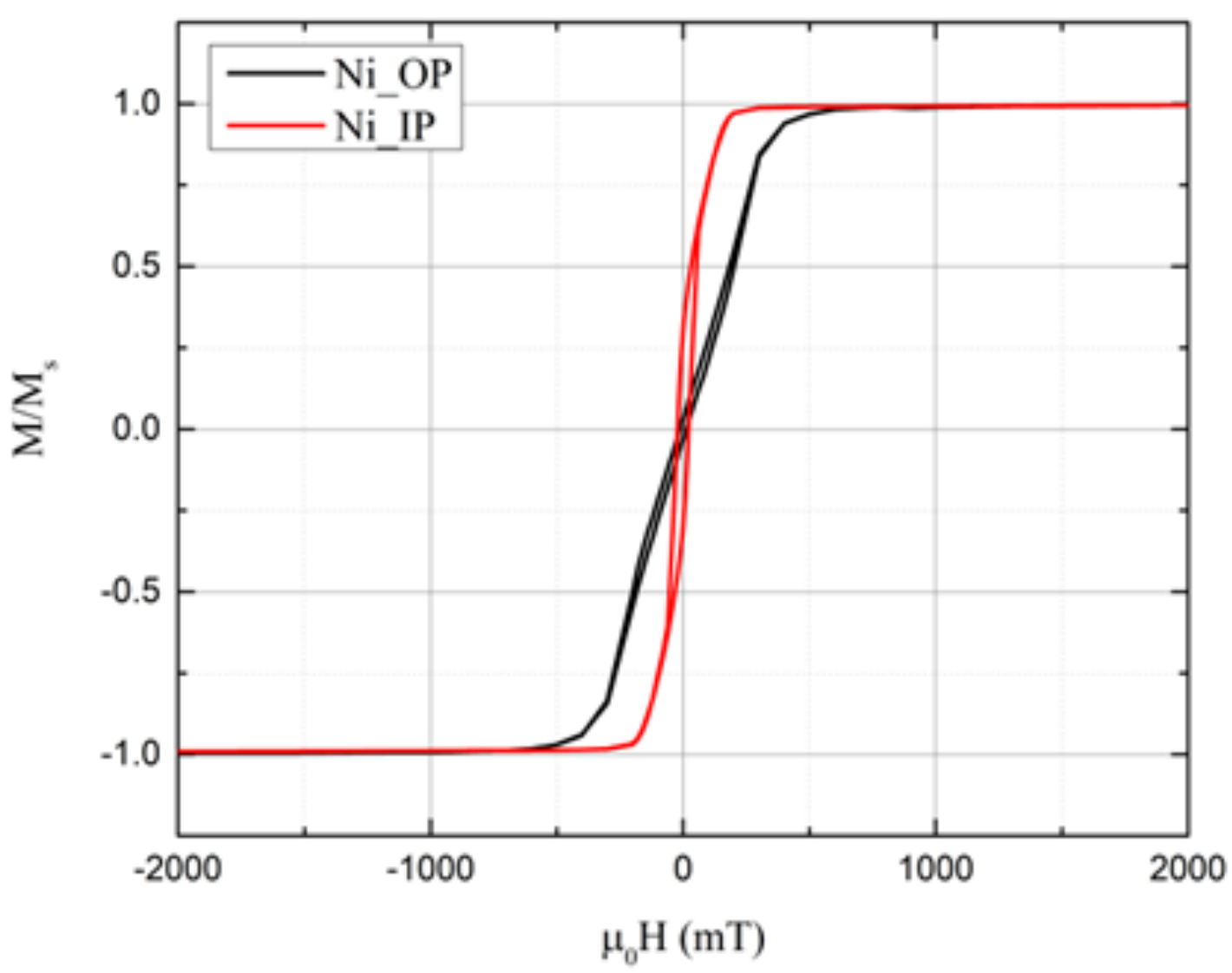

**Figure S13:** Magnetization curves of 100 nm Ni thin films measured at 300 K, showing IP (Ni_IP, red) and OOP (Ni_OP, black) magnetization as a function of the applied magnetic field.

To ensure that the magnetic behavior is preserved after the annealing step in the SAM deposition procedure, SQUID magnetization measurements were performed both before and after thermal treatment. The substrate was annealed on a hot plate at 100 °C for 1 hour, after which SQUID measurements were repeated. As shown in Fig. S14, the magnetization curve remains unchanged before and after the heat treatment. The saturation field remains approximately at ± 800 mT.

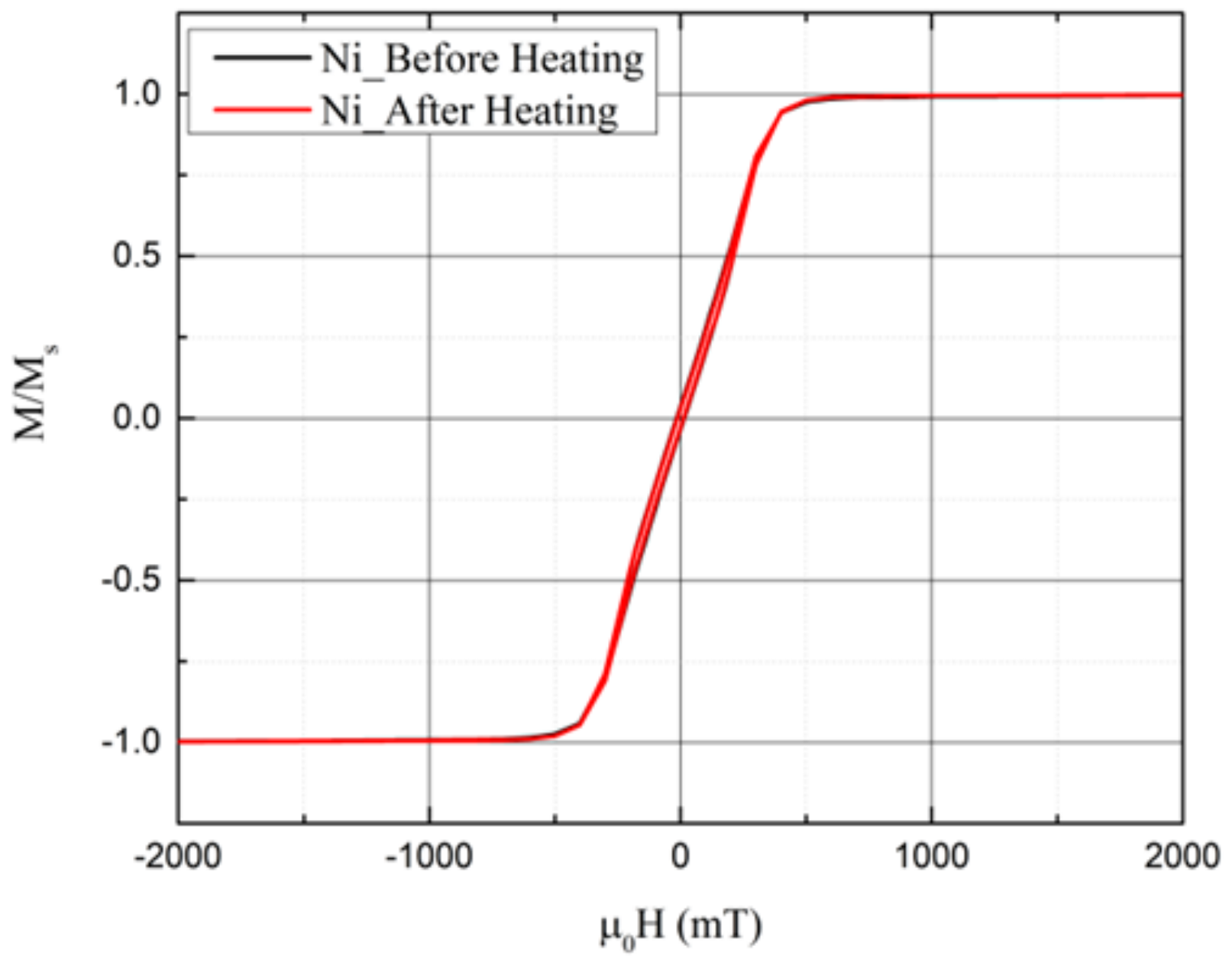

**Figure S14:** Magnetization as a function of the applied magnetic field of 100 nm Ni thin films measured before heating (black) and after heating at 100 °C for 1 hour (red).